\documentclass{pasj01}
\usepackage[switch,mathlines]{lineno}

\usepackage{longtable}
\Received{$\langle$reception date$\rangle$}
\Accepted{$\langle$acception date$\rangle$}
\Published{$\langle$publication date$\rangle$}

\begin{document}

\title{X-ray Timing and Spectral studies of bare AGN Mrk 110}

\author{Deblina Lahiri$^{1}$}
\altaffiltext{1}{Department of Astronomy, University College of Science, Osmania University, Hyderabad 500007, India}
\altaffiltext{2}{Space Astronomy Group, ISITE Campus, U. R. Rao Satellite Centre, Outer Ring Road, Marathahalli, Bangalore - 560037, India.}

\author{K. Sriram$^{1}$}

\author{Vivek K. Agrawal$^{2}$}

\email{lahiri.deblina1997@gmail.com}


\maketitle

\begin{abstract}
The origin of the soft X-ray excess below $\sim$2 keV in active galactic nuclei (AGNs) remains debated, with relativistic reflection from the inner accretion disk and warm Comptonization in an optically thick corona being the leading explanations. We investigate the timing and spectral properties of the Seyfert galaxy Mrk 110 using six XMM–Newton observations. A frequency–dependent lag analysis in the 7 -- 9 $\times 10^{-5}$ Hz range reveals a soft X-ray lag of $\tau_{\rm soft} \sim$ $-$889 -- $-$3000 s in the combined 2019 data, detected with a significance of $\sim$80\%. The cross-correlation function analysis, supported by simulations, also detects lags of similar nature. Spectral modeling performed by adopting both proposed black hole masses in the literature for Mrk 110 confirms the presence of a warm corona in all observations, along with a weak relativistic reflection component and the reflection fraction remains low (R$_{\rm f}$ $<$ 1). Interpreting the measured soft lag in terms of light travel time implies an emission radius $\sim$ 4.5 R$_{\rm g}$ for a supermassive black hole mass of M = 1.4 $\times$ 10$^{8}$ M$_{\odot}$, favoring a reflection scenario. However, if a lower mass of M = 2 $\times$ 10$^{7}$ M$_{\odot}$ is adopted, the inferred radius increases, and both relativistic reflection and warm Comptonization can plausibly contribute to the observed soft lag. The warm corona radius appears larger in the high accretion state and smaller in a lower accretion state, although no trend can be established. The persistently low reflection fraction suggests an outflowing inner corona in Mrk 110, consistent with the recent detection of jet activity in this source.

\end{abstract}

\linenumbers
\section{Introduction}

 Active galactic nuclei (AGNs) are compact regions in galaxies that have supermassive black holes (SMBHs) at their center along with an accreting disk of gas/dust that is optically thick and geometrically thin in nature (Shakura \& Sunyaev 1973) along with a Comptonization region i. e. corona. The location and structure of the corona are still a matter of debate viz. radial and vertical dimensions of the corona, its connection to the jet, its effect on inner and outer regions of the disk, etc. Much of the properties of the corona and its effects on the Keplerian portion of the disk can be understood by studying the X-ray temporal and spectral data. 

{\bf X-ray variability studies have revealed the presence of soft lags i. e. soft photons lag the continuum hard X-ray photons showing negative lag in the frequency-lag spectrum {\bf around 10$^{-5}$ -- 10$^{-3}$ Hz} and have been seen in many of the AGN sources (De Marco et al. 2013; Kara et al. 2016). Soft lags can be naturally interpreted within the standard X-ray reflection paradigm (Guilbert \& Rees 1988; Fabian et al. 2009). In this scenario, the primary X-ray continuum is produced in a compact corona, while a fraction of the emission irradiates the accretion disk and is reprocessed into a reflection spectrum composed of a continuum and emission lines (e.g. Ross \& Fabian 2005). Reflection originating from the inner disk is strongly modified by Doppler shifts and relativistic effects in the vicinity of the black hole. The resulting soft lags, where fluctuations in the primary continuum precede the soft reflected response, are attributed to the light travel time delay between the corona and the inner accretion disk.

A key advantage of the relativistic reflection model is its ability to naturally account for the soft excess observed in the X-ray spectra of active galactic nuclei (Fabian \& Ross 2010). Another leading physical interpretations is 
 the warm corona model which attributes the soft excess to thermal Comptonization of disk photons in a moderately hot (kT $\sim$0.1 -- 0.5 keV), optically thick plasma ($\tau$$\sim$ 10 -- 20) that extends over the inner accretion disk (Petrucci et al. 2018). Both models can successfully reproduce the smooth spectral shape of the soft excess and are often difficult to distinguish using spectral fitting alone, motivating the use of timing diagnostics such as energy-dependent time lags to disentangle their relative contributions. Soft lags frequently support the reflection interpretation in AGNs, although alternative models invoke different physical mechanisms in a few AGNs (eg. Miller et al. 2010, Zoghbi et al. 2011; Gardner \& Done 2014; Silva et al. 2016; Sriram et al. 2023).}

{\bf In several Seyfert galaxies, the X-ray spectra are successfully described within a two-corona model, in which the emission arises from physically distinct warm and hot coronal regions. In this scenario, the soft X-ray emission is modeled through Comptonization in a warm corona using components such as \texttt{CompTT} (Titarchuk 1994) or \texttt{nthComp} (Zdziarski et al. 1996), while the hard X-ray continuum is attributed to a hot corona and is commonly represented by a power-law or \texttt{CompPS} (Poutanen \& Svensson 1996). This approach has been successfully applied to sources such as Mrk 509 (Petrucci et al. 2013), RX J0439.6–5311 (Jin, Done \& Ward 2017), samples of AGNs (Petrucci et al. 2018), NGC 7469 (Middei et al. 2018), NGC 4593 (Middei et al. 2019), HE 1143–1810 (Ursini et al. 2019), and Mrk 359 (Middei et al. 2020).
The AGNSED model provides a physically self consistent description of the AGN broadband spectral energy distribution by coupling thermal emission from the accretion disk with Comptonization in warm and hot coronae. It conserves the mass accretion rate and energy budget, enabling the coronal geometry and spectral properties to be constrained within a unified framework (Kubota \& Done 2018).}

The AGN Mrk 110 has a low redshift $z$ = 0.035291 and is classified as a broad line Seyfert (BLS 1) as it exhibits a large [O III] 5007/H$\beta$ ratio (Grupe 2004; Véron-Cetty et al. 2007). Initially, it was classified as NLS1 because only the narrow component of the H$\beta$ line emitting from the BLR was considered (Boller et al. 2007; Gliozzi \& Williams 2020). {\bf However, more detailed studies reveal that the H$\beta$ emission lines, together with the He I and He II lines, exhibit broader profiles with their centroids systematically shifted toward longer wavelengths (Bischoff \& Kollatschny 1999; Véron-Cetty et al. 2007) that consistently classify Mrk 110 as a BLS1 source. Reverberation mapping studies yield masses of 1.8--2.5 $\times$ 10$^{7}$ M$_{\odot}$ (Kollatschny et al. 2001; Peterson et al. 2004; Du \& Wang 2019), consistent with estimates from the bulge stellar velocity dispersion (Onken et al. 2004). By contrast, interpreting the systematic redshift of the broad emission lines as gravitational redshift, Kollatschny (2003) derived a substantially higher mass of M = 1.4 $\times$ 10$^{8}$ M$_{\odot}$, in agreement with spectro-polarimetric measurements of 2.1$\pm$0.4 $\times$ 10$^{8}$ M$_{\odot}$ (Afanasiev et al. 2019).} Mrk 110 is a radio quiet (RQ) source exhibiting short time scale variability in GHz providing a spacial limit of emitting region of 180 $R_{\rm s}$ (Panessa et al. 2022). {\bf However, the discovery of a relativistic jet in Mrk 110, previously considered as a radio quiet AGN is a recent finding (Wang et al. 2025). The VLBI observations reveal intermittent jet activity during 2015–2016 and 2022–2024, with {\bf apparent} superluminal motions of $\sim$2.1 -- 3.6c. This positions Mrk 110 as a crucial missing link between radio-loud and radio-quiet AGN, offering new insights into jet launching and unification scenarios.} 

{\bf Our study aims to robust the detection of soft lags using the publicly available \textit{XMM-Newton} data, thereby increasing the total exposure time by a factor of $\sim$7 compared to that analyzed by De Marco et al. (2013) and to further test warm Comptonization models as the origin of the soft excess and search for possible evolution in coronal parameters across these \textit{XMM-Newton} observations of Mrk 110.}

The structure of the paper is as follows. Observations and data reduction are discussed in Section 2 {\bf and detection and reliability of soft lag through a detailed timing analysis is described in Section 3. In section 4, extending the analysis of Porquet et al. (2024), we employ various models to provide a more detailed interpretation of the broadband UV–X-ray spectrum of the six observation obtained from the \textit{XMM-Newton} satellite.} Our discussion and conclusions are given in Sections 5 and 6, respectively.

\section{Observations and Data Reduction}
 The {\it XMM-Newton} archival data of the AGN source Mrk 110 is used to investigate temporal and spectral variations. The first observation (Obs. 1: Obs. Id. 0201130501) was carried out on 2004-11-15 with a duration of 47.42 ks. 
 For the second, 2019 observations and its timing analysis, we have considered four sets of observations that were made a few days apart (Obs. 2a: Obs. Ids. 0840220701), (Obs. 2b: Obs. Ids.  0840220801), (Obs. 2c: Obs. Ids. 0840220901) and (Obs. 2d: Obs. Ids. 0852590101) taken on 2019-11-03, 2019-11-05, 2019-11-07 and 2019-11-17 with exposure time 43.6 ks, 43 ks, 40.6 ks and 44.5 ks respectively. From here onward, the combined results of all four observations of 2019 will be referred as Obs. 2.
The third observation (Obs. 3: Obs. Id. 0852590201) was performed on 2020-04-06 for 48.5 ks. 
We have taken the EPIC-pn data for the timing analysis, as it has a better signal-to-noise ratio compared to the other cameras onboard.  For spectral analysis, data from all three EPIC instruments viz. pn, MOS1 and MOS2  were used to improve spectral coverage and statistics.
For all the above mentioned datasets EPIC-pn camera used a thin filter with prime small window mode,  while the MOS1 \& MOS2 cameras used a thin filter in full-frame mode.
We have utilized the \textit{XMM-Newton} Science Analysis Software (SAS) version 19.1.0 and the current calibration files (CCF).
For extraction of the source and a background (source-free) region to obtain the light curves and spectra, we have used the {\it evselect} task
 with {\it PATTERN $\le$ 4} for pn and {\it PATTERN $\le$ 12} for MOS1/MOS2 along with the {\it epiclccorr} task to get the background corrected light curves in the desired energy bands for the present study.
 Here, we have used the {\it epatplot} task to check for the pile-up effect associated with the data sets. {\bf Pile-up effects were determined in MOS1 and MOS2 of Obs. 1, MOS1 of Obs. 2c and MOS1 of Obs. 2d therefore the source spectrum was extracted from an annular region with inner and outer radii of 10$^{''}$ \& 40$^{''}$ , respectively for these datasets.}
 The observations other than those mentioned shows no pileup effect, and we considered a circular region of radius 40$^{''}$ for the source and the respective background regions. Figure 1 displays the X-ray light curves from EPIC-pn camera with a bin time t = 500 s for all the observations.
 For each instrument, the SAS tasks {\it rmfgen} \& {\it arfgen} were used to get the redistribution matrix file (RMF) and ancillary response file (ARF) respectively. We fitted the broad band spectra and used {\it XSPEC} v12.12.0 (Arnaud 1996) for all the spectral analysis carried out in this work. The $\chi^{2}$ statistics is used to get the best-fit spectral parameters and the uncertainties are reported at a 90\% confidence level.

 Optical Monitor (XMM-OM, Mason et al. 2001) provides simultaneous optical-UV photometry for all XMM-Newton observations for the source Mrk 110. A total of three filters were employed: UVW1 ($\lambda_{eff}$ = 291 nm), UVM2 ($\lambda_{eff}$ = 231 nm), UVW2 ($\lambda_{eff}$ = 212 nm). The concurrent optical-to-X-ray coverage offers crucial details for comprehending how the disk interacts with its surroundings (BLR and corona). Running the SAS job {\it omichain} with default input parameters resulted in the reduction of XMM-OM data. In every observation, we also used {\it om2pha} to extract the spectral points for each filter. A Galactic reddening of E(B - V) = 0.01 for Mrk 110 was considered (Schlafly \& Finkbeiner 2011) and the  distance
is fixed to 155.7 Mpc (Wright 2006; Planck Collaboration VI 2020) and we added a systematic of 1.5\% to the UV spectra (Porquet et al. 2024).

\begin{figure}
    \centering
    \includegraphics[height=\linewidth, angle=270]{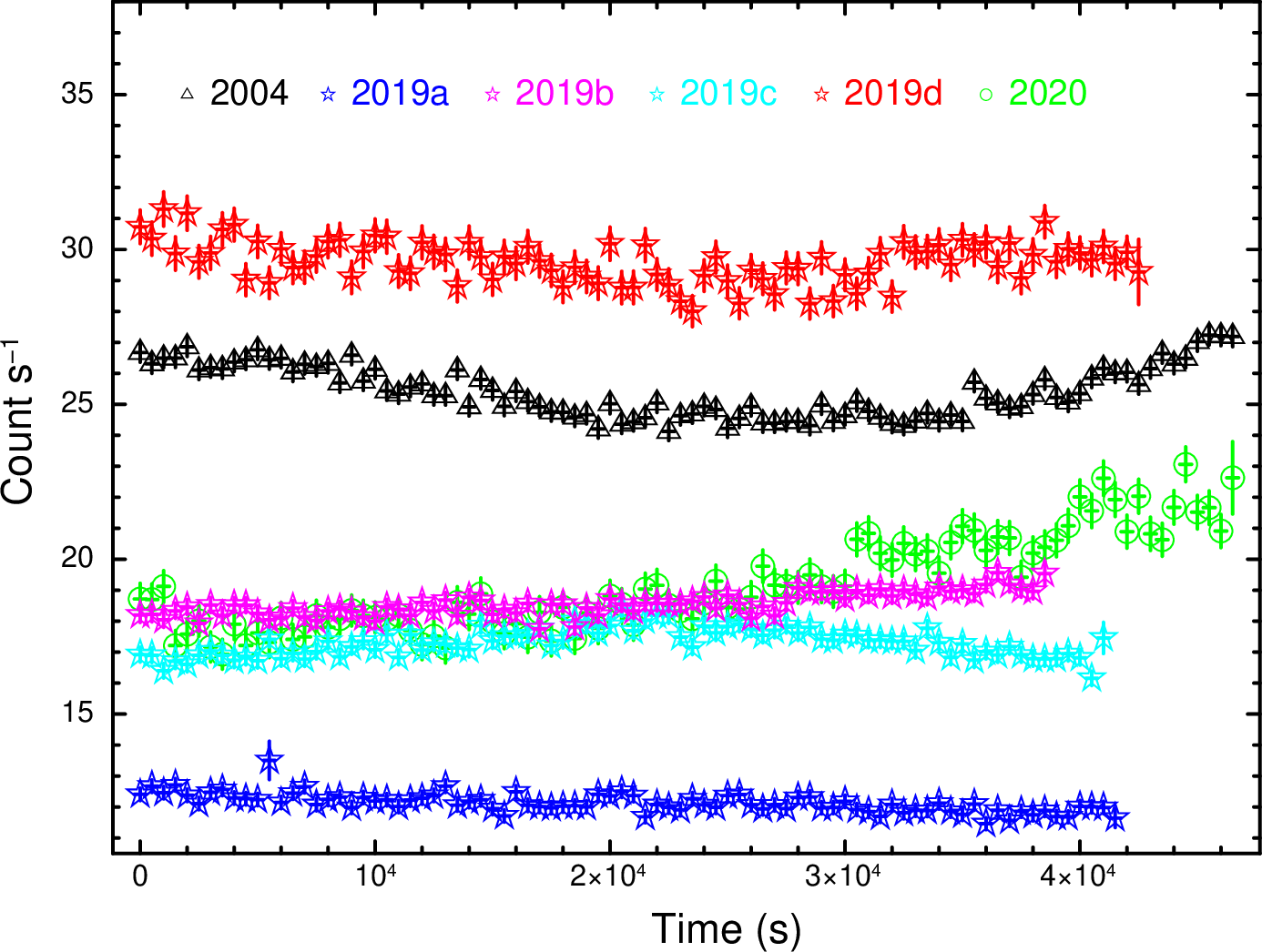}
        \caption{ EPIC-pn X-ray light curves for all observations in 0.3 -- 10 keV energy bands with a time bin of 500 s. ///
        \\Alt text:Plot showing all the x-ray light curve for the source}
    \label{fig:enter-label}
\end{figure}
\begin{figure}
    \centering
    \includegraphics[height=\linewidth, angle=270]{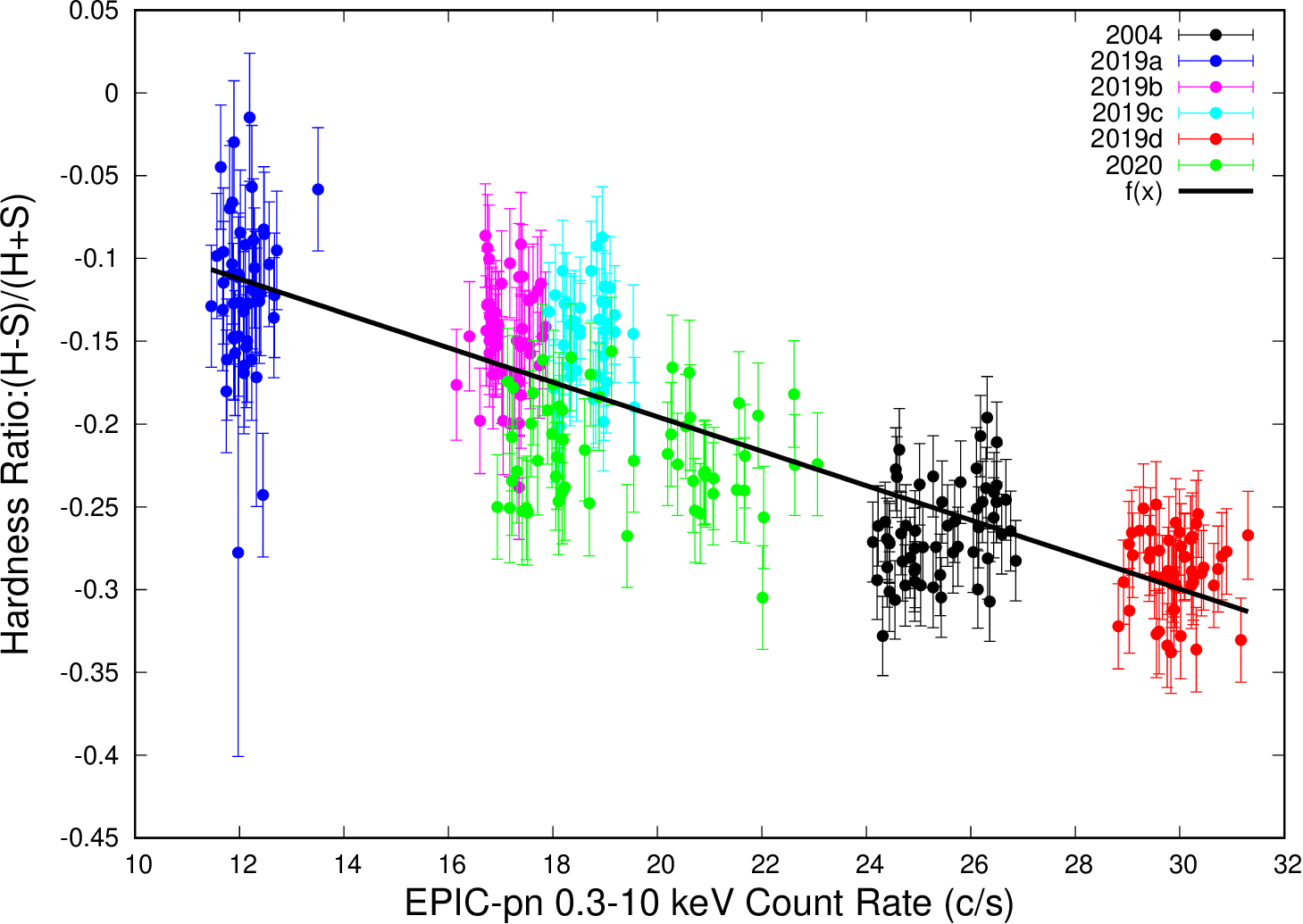}
        \caption{ Hardness-intensity plot for all the observations along with a linear fit (f(x)), see text for more details.///
        \\Alt text:Plot of  all the hardness ratio along with the fit in shown.}
    \label{fig:enter-label}
\end{figure}

\section{Timing Analysis}

The long-term variability of the source was inspected by computing the hardness ratio for all observations (Fig. 2). We extracted the soft (S, 0.3 -- 1 keV) and hard (H, 1 -- 5 keV) band light curves and calculated the hardness ratio as $\frac{H-S}{H+S}$. Later the ratio was plotted against the total 0.3--10 keV count rate. It was observed that the source was soft when brighter which is often seen in other radio-quiet Seyferts (eg. Lobban et al. 2018). We fitted a linear function i. e. y = m*x+c and obtained a best-fit with a slope m = $-$0.016 $\pm$ 0.001 and c = 0.02 $\pm$ 0.01 with a fit statistic of $\chi^{2}$/dof = 489/304 (Fig. 2). To check the consistency of the fit as the source displays different flux variability from 2004 to 2020, we removed the data contributing to the high hardness ratio but still noticed a similar linear trend with no significant change in the slope.

\subsection{Power Density Spectrum}
The power density spectrum (PSD) was extracted to examine the strength of variability in the frequency domain. The light curves for each observation of Mrk 110 with a time resolution of $\Delta$ t = 100 s were used to calculate the periodograms (Vaughan et al. 2003) and estimate the PSD (Fig. 3). 

We modeled the raw periodograms in XSPEC using a simple power–law plus constant model (Lobban et al. 2018), P ($\nu$) = N $\times$ $\nu^{-\alpha}$ + C,
where $\alpha$ is the spectral index, 
N is the power–law normalization, and 
C represents a flat component accounting for the high-frequency Poisson noise. The latter becomes dominant at frequencies above $\sim$ 1-2 $\times$ 10$^{-4}$ Hz. Each epoch of 2004, 2019, 2020 was fitted individually. 

The best-fitting parameters were obtained by minimizing the Whittle statistic (see for more details; Vaughan 2010; Lobban et al. 2018), which is particularly well suited for fitting power spectral densities. We noted that $\alpha$ = 1.86$^{+0.83}_{-0.71}$ with Whittle(S)=$-$1244 for the lowest flux observation Obs. 2a and $\alpha$ = 2.07$^{+0.96}_{-0.64}$ with Whittle (S) = $-$1359 for the highest flux state Obs. 2d. The slopes of other observations were found to vary in between the above mentioned values (see Fig. 3) and these slopes are used in the Monte-Carlo simulations that we have carried out later for testing the lag significance.



\begin{figure}
    \centering
    \includegraphics[height=\linewidth, angle=270]{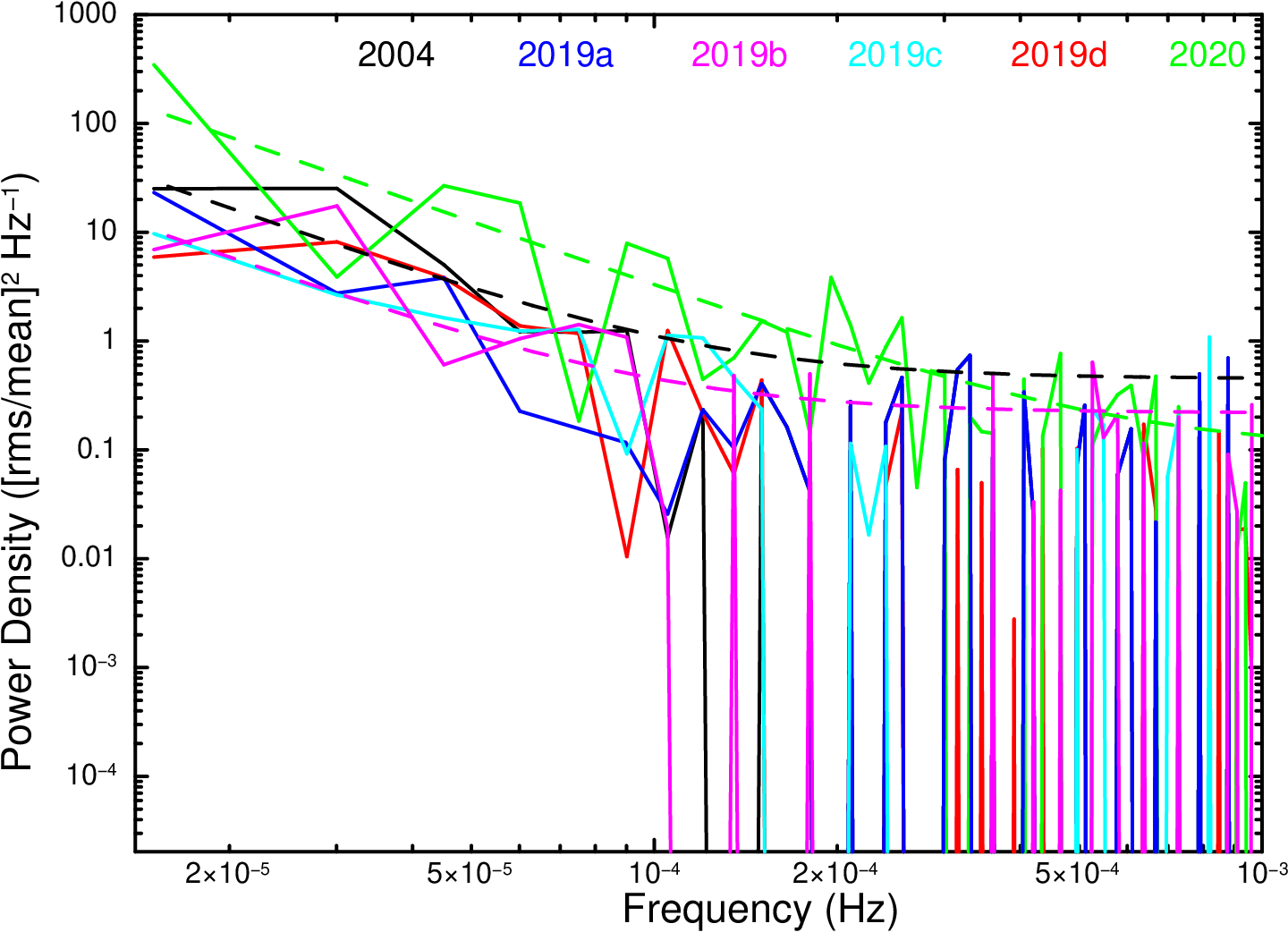}
        \caption{Power density spectra in 0.3 -- 10 keV energy band of Mrk 110 for different observations. The dotted line shows the best fit powerlaw+constant model for three of the observations with the same color as the dataset for clarity (Obs. 1, Obs. 2b, and Obs. 3).///
        \\Alt text: All the power density spectra are over-plotted along with the fitting}
    \label{fig:enter-label}
\end{figure}

\subsection{Frequency-lag and Coherence}
We measure the frequency-dependent time lags between the soft (0.3 -- 1 keV) and hard (1 -- 5 keV) X-ray bands by evaluating the lag as a function of temporal frequency. The lags are computed using the Fourier method described in Nowak et al. (1999) and Uttley et al. (2014). Two evenly sampled light curves, s(t) and h(t) corresponding to the soft and hard bands, are transformed into the frequency domain to obtain their Fourier transforms

S($\nu$) and H($\nu$) where S($\nu$)= S($\nu$) e$^{i{\phi_s(\nu)}}$, and H($\nu$)= H($\nu$) e$^{i{\phi_h(\nu)}}$. 

The phase difference between the two bands is derived from the cross-spectrum \\
\begin{equation}C(\nu)= S^{*}(\nu)H(\nu)= S(\nu) H(\nu) e^{i[{\phi_h(\nu)}-{\phi_s(\nu)}]}\\  
\end{equation}


from which the frequency-dependent time lag is obtained: \\
\begin{equation}
  \tau(\nu)=[{\phi_h(\nu)}-{\phi_s(\nu)}] / 2\pi\nu  
\end{equation}

Negative lags indicate that the soft band lags behind the hard band (soft lag), whereas positive lags correspond to hard lags.
Figure 4 (top panels) shows the lag spectra for Obs. 1 (2004; black triangles), Obs. 2 (combined 2019; red stars), and Obs. 3 (2020, green circles). We noted a soft lag of $\tau$ = $-$3067 $\pm$ 1412 s at 9 $\times$ 10$^{-5}$ Hz and $-$889 $\pm$ 412 s at 7 $\times$ 10$^{-5}$ Hz (Obs. 2, 2019 all combined). In 2004, the soft lag was observed to be around $\tau$ = $-$401 $\pm$ 468 s at 9 $\times$ 10$^{-5}$ Hz, whereas no such soft lag was observed in Obs. 3. {\bf After combining the 2019 and 2020 observations, we find that both the soft and hard lags are reduced in amplitude, likely due to the lower intrinsic variability to noise ratio of the source.} Coherence was calculated for the light curves between the soft energy and the hard energy bands (Vaughan \& Nowak 1997; Nowak et al. 1999). Figure 4 shows that the coherence remains moderate at low frequencies ($\sim$ 0.51 $\pm$ 0.14), but become increasingly incoherent at higher frequencies as Poisson noise begins to dominate.

 {\bf To validate the significance of soft lag, we used the combined observation of 2019 i. e. Obs. 2}  adopting the methodology used by De Marco et al. (2013). Following the Timmer \& König (1995) method, 5000 pairs of simulated light curves in the 0.3 -- 1 keV and 1 -- 5 keV bands were produced using the power-law slopes derived from the PDS fitting in the present work, along with the corresponding observed means and standard deviations. For these pair of simulated light curves, we calculated frequency-lag spectra with similar constrains used for real frequency-lag spectra and {\bf it must be noted that we have not introduced Poisson noise as it is less in low frequency domain and the pair of simulated light curves had zero intrinsic lag.} To obtain the significance of lags, a sliding frequency window is defined (as N$_{ \rm w}$=1) in the frequency domain containing soft lags. The merit of lags was estimated by calculating $\chi$ = $\sqrt{\Sigma (\tau / \sigma_{\tau})^2} $
in the frequency domain where Poisson noise is not dominant. We obtained all the $\chi$ from the simulated data which were more than the observed $\chi$ obtained from the real frequency-lag spectra and this limit would be due to the occurrence of lags by chance. {\bf For the combined data of Obs. 2 (a, b, c, d), we obtain an upper limit significance of 80\% for the observed soft lag.} The plot between $\chi$ and the number of simulations (N) is shown in Figure 5 where the red dotted line represents the real value $\chi$ obtained from the observed frequency lag spectra.  \\

\begin{figure}
    \centering
    \includegraphics[height=8cm, width=15cm,, angle=270]{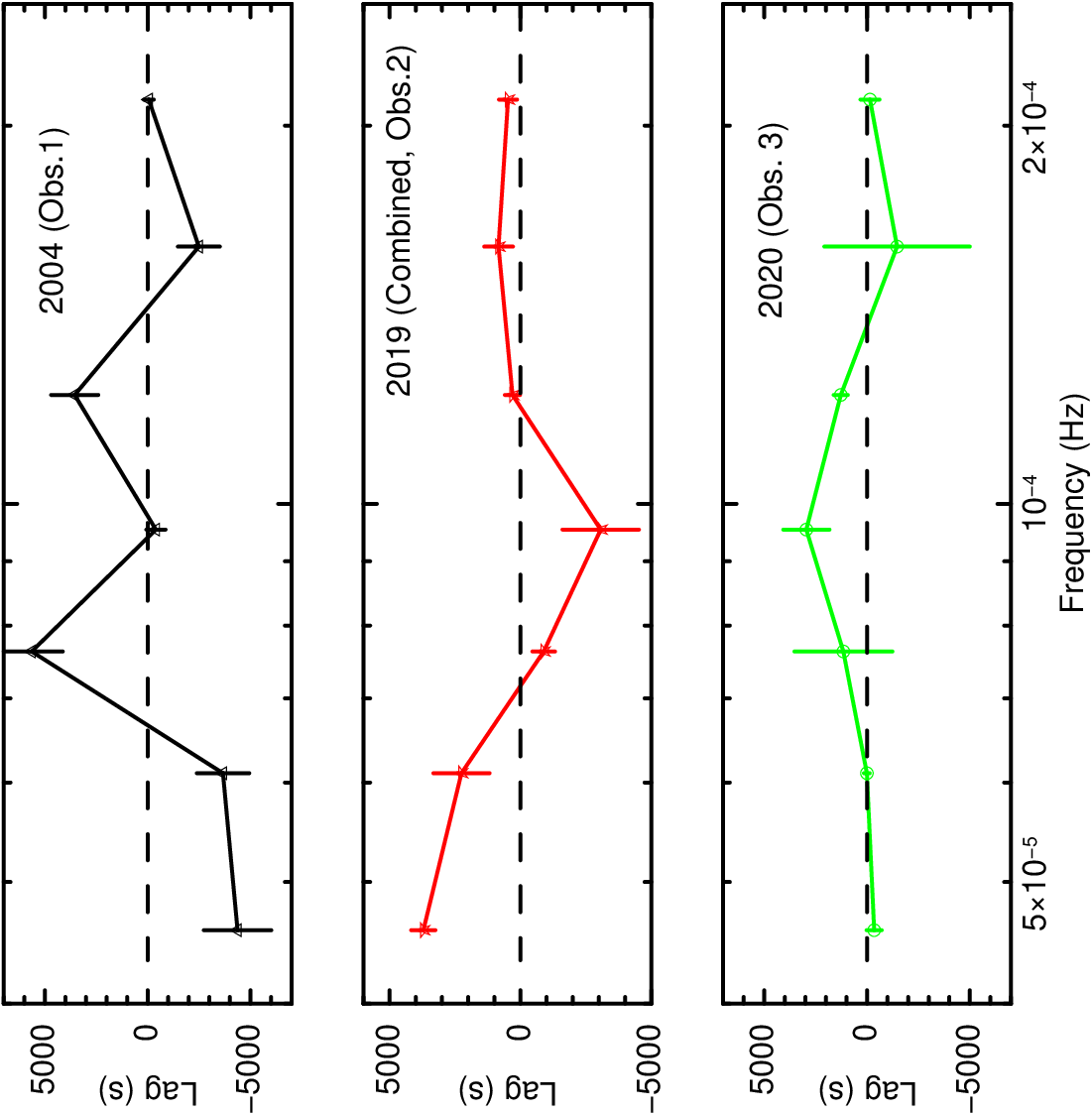}
    \includegraphics[height=8cm, width=5cm, angle=270]{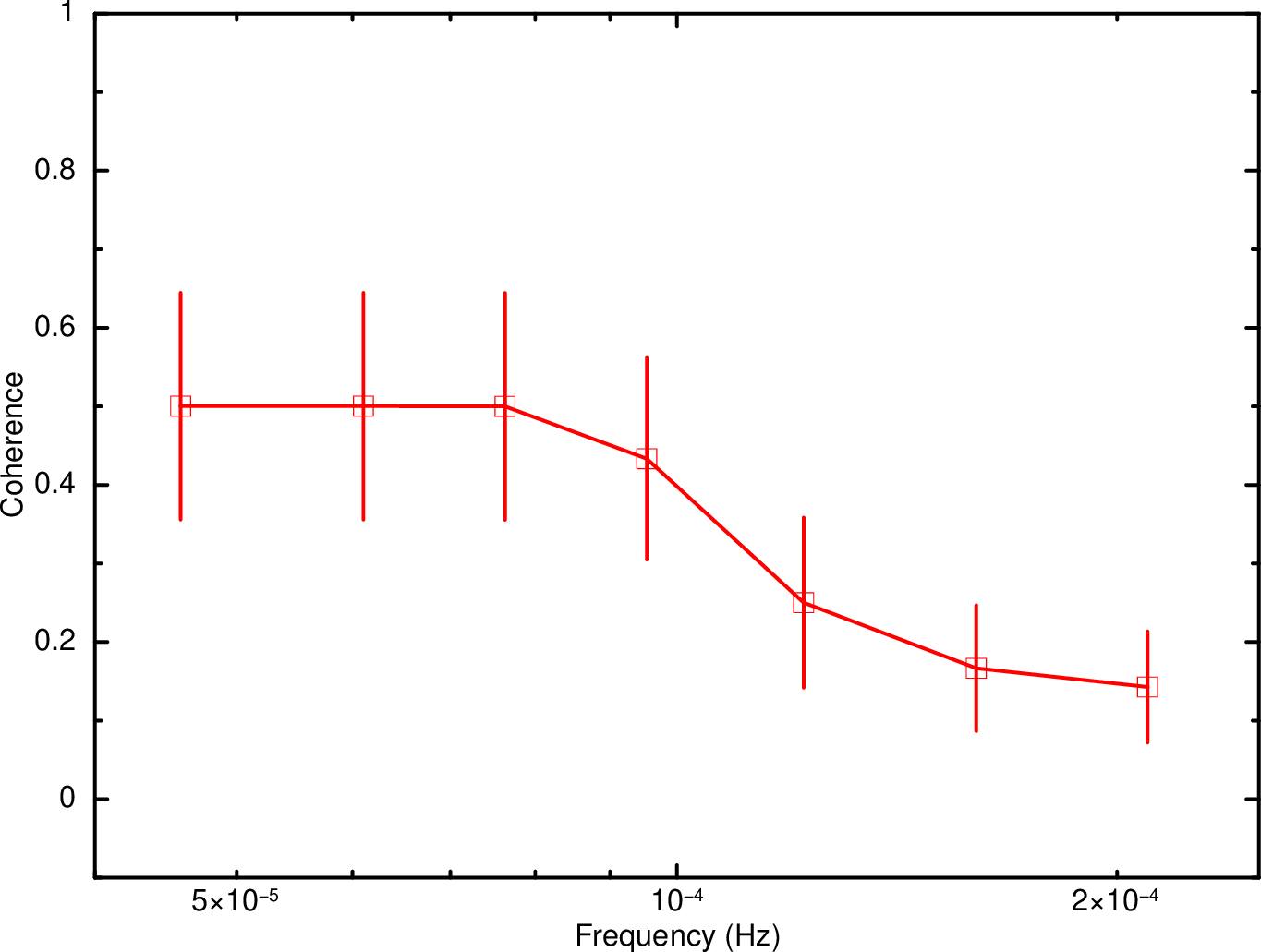}\\
    \caption{ Top three panel displays the frequency-lag spectra (0.3 -- 1 keV versus 1 -- 5 keV) of the respective observations. The dashed line displays the zero lag. The bottom panel displays the coherence with 1$\sigma$ error bars.///
    \\Alt text: The top three panels shows the frequency-lag spectra and the bottom panel shows the coherence plot for all the observations. } 
    \label{fig:enter-label}
\end{figure}

\begin{figure}
    \centering
    \includegraphics[width=\linewidth, angle=0]{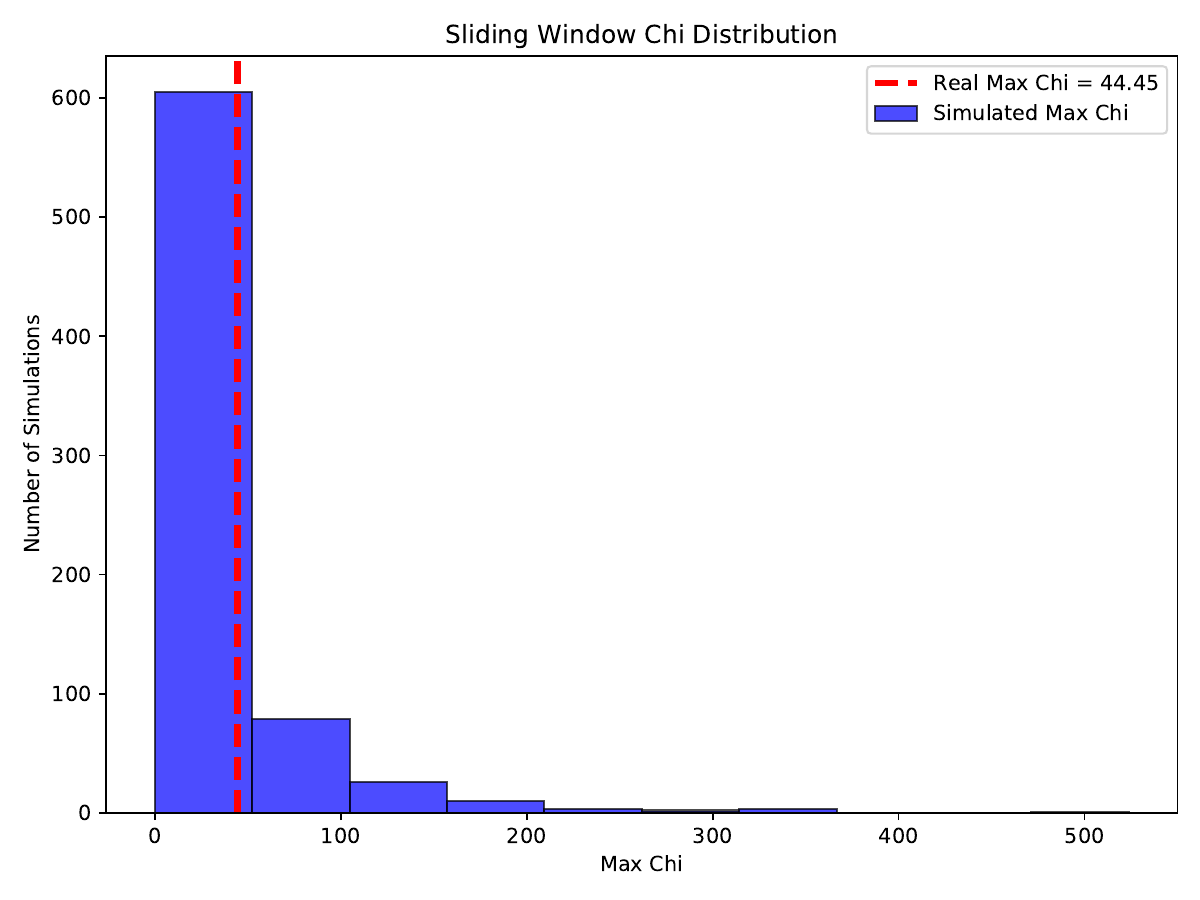}
    \caption{ The merit of lag $\chi$ for all the simulated frequency-lag spectra for all the 2019 observation combined  to obtain the soft-lag significance is plotted. The red-dotted line represents real $\chi$ value of the observation.///
    \\Alt text:  Histogram plot for the merit of lag and the real merit value}
    \label{fig:enter-label}
\end{figure}

\subsection{Energy-lag Spectrum}
The energy-lag spectra for the observations were computed using the standard procedure elaborately described in Uttley et al. (2014). We have selected a broader energy bin for the reference band (0.3 -- 10 keV) and few small energy bins or the {\it channel-of-interest} (CI). We have subtracted the CI from the reference band and obtained a CI-corrected reference light curve that is used for calculating the cross spectrum to obtain the lag spectra. It provides good signal to noise ratio as the error associated with Poisson noise is minimized. Figure 6 displays the respective energy-lag spectra of different observations from left to right panels at frequency 9 $\times$10$^{-5}$ Hz. 
  

\begin{figure}
    \centering
    \includegraphics[height=8cm, width=10cm,, angle=270]{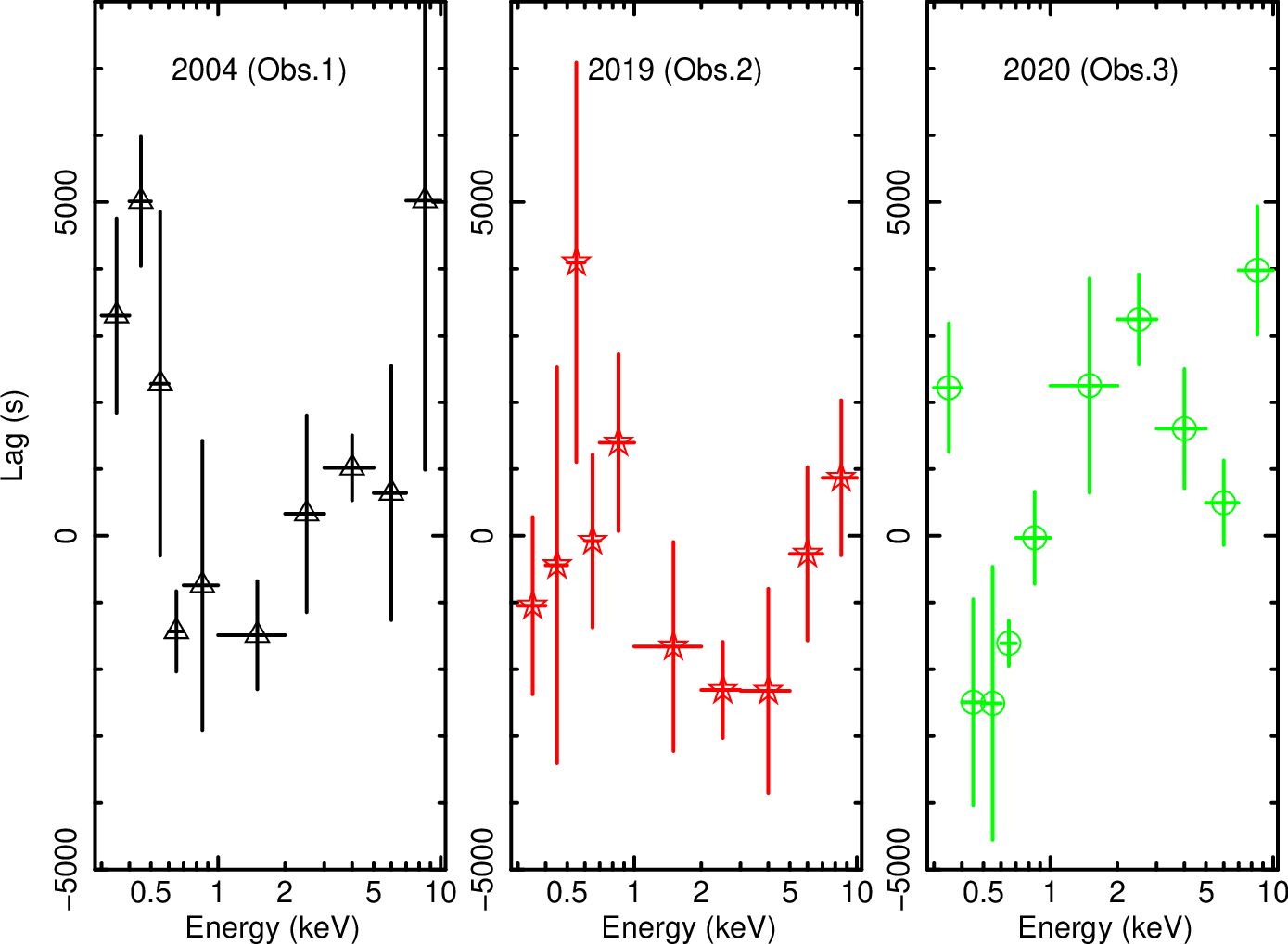}
 \caption{The panels from left to right shows energy-lag spectra of Obs. 1 (black-triangle symbol), Obs. 2 (red-star symbol) and Obs. 3 (green-circle symbol) of Mrk 110 at 9 $\times$10$^{-5}$ Hz. ///
 \\Alt text:The energy lag plot for all the observations }
    \label{fig:enter-label}
\end{figure}

\subsection{X-ray Cross Correlation Function}
The study of cross-correlation function (CCF) also provides an evidence of soft lags in AGNs (eg. Ark 120, Nandi et al. 2021). To confirm soft lags, we carried out cross-correlation function (CCF) analysis between 0.3 -- 1 keV and 1 -- 5 keV for Mrk 110. To verify the significance, we simulated CCFs using light curves produced with the Timmer \& König (1995) method, adopting the PSD slope, mean, and standard deviation of the observed light curves, with a 1000 s bin size. 
Figure 7 displays all the simulated CCFs (blue shaded region), observed CCF (red line), and 95\% confidence plot (dashed line) for Obs. 1 (2004) followed by Obs. 2a, 2b, 2c, \& 2d (2019) along with Obs. 3 (2020). The right panels display the fitted lag histogram for the respective observations. For Obs. 1 and Obs. 2b, we noted a slightly asymmetric CCF peaking towards hard lag (
$\tau_{\rm ccf}$ $\sim$ 2000 s), in Obs. 2a, 2c \& 2d a soft lag ($\tau_{ccf}$ $\sim$ $-$1018 -- $-$1621 s) is relatively prominent and in Obs. 3 the CCF was highly symmetric.

\begin{figure*}
\includegraphics[height=3 cm, width=8.5cm, angle=0]{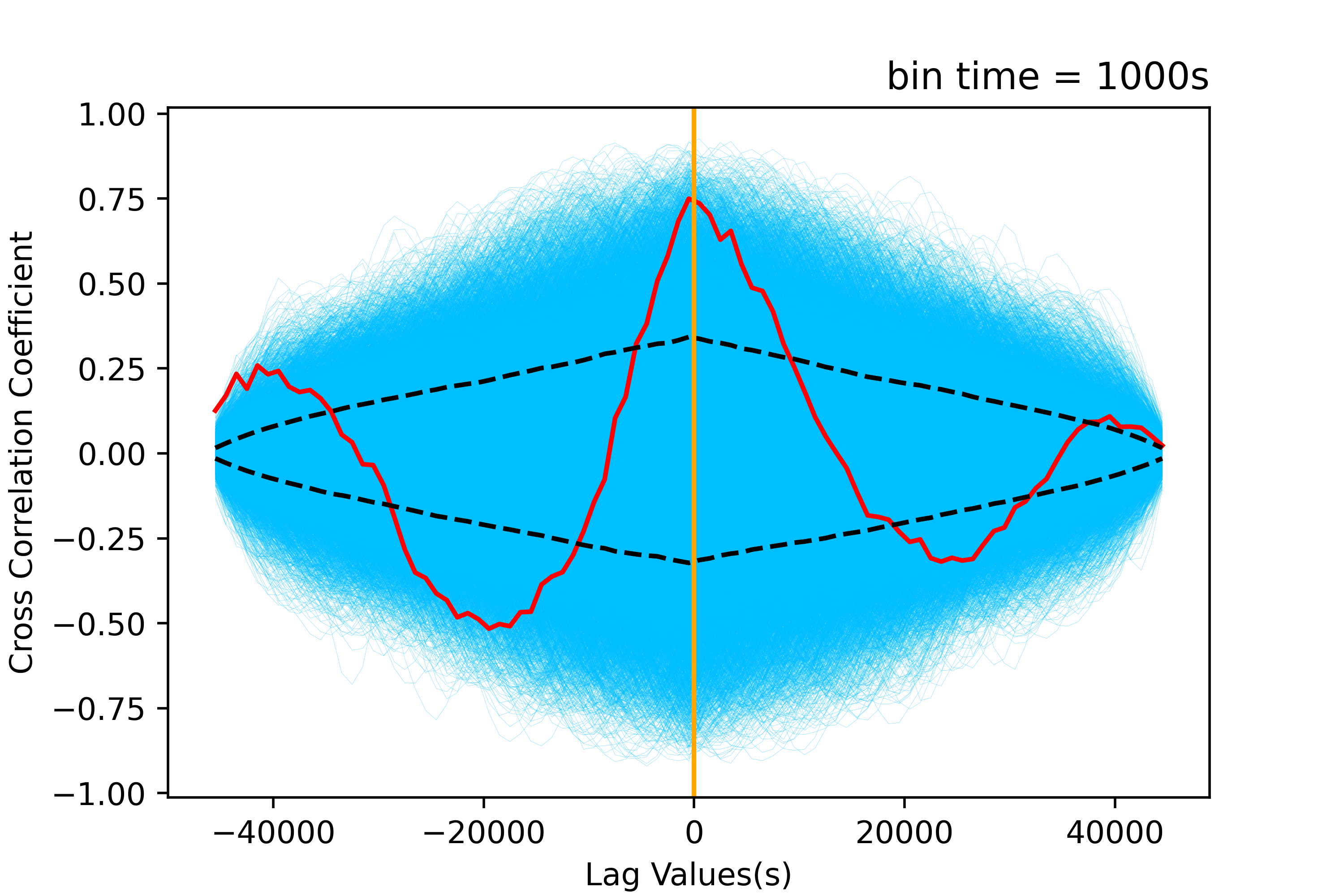} 
\includegraphics[height=3 cm, width=8.5cm, angle=0]{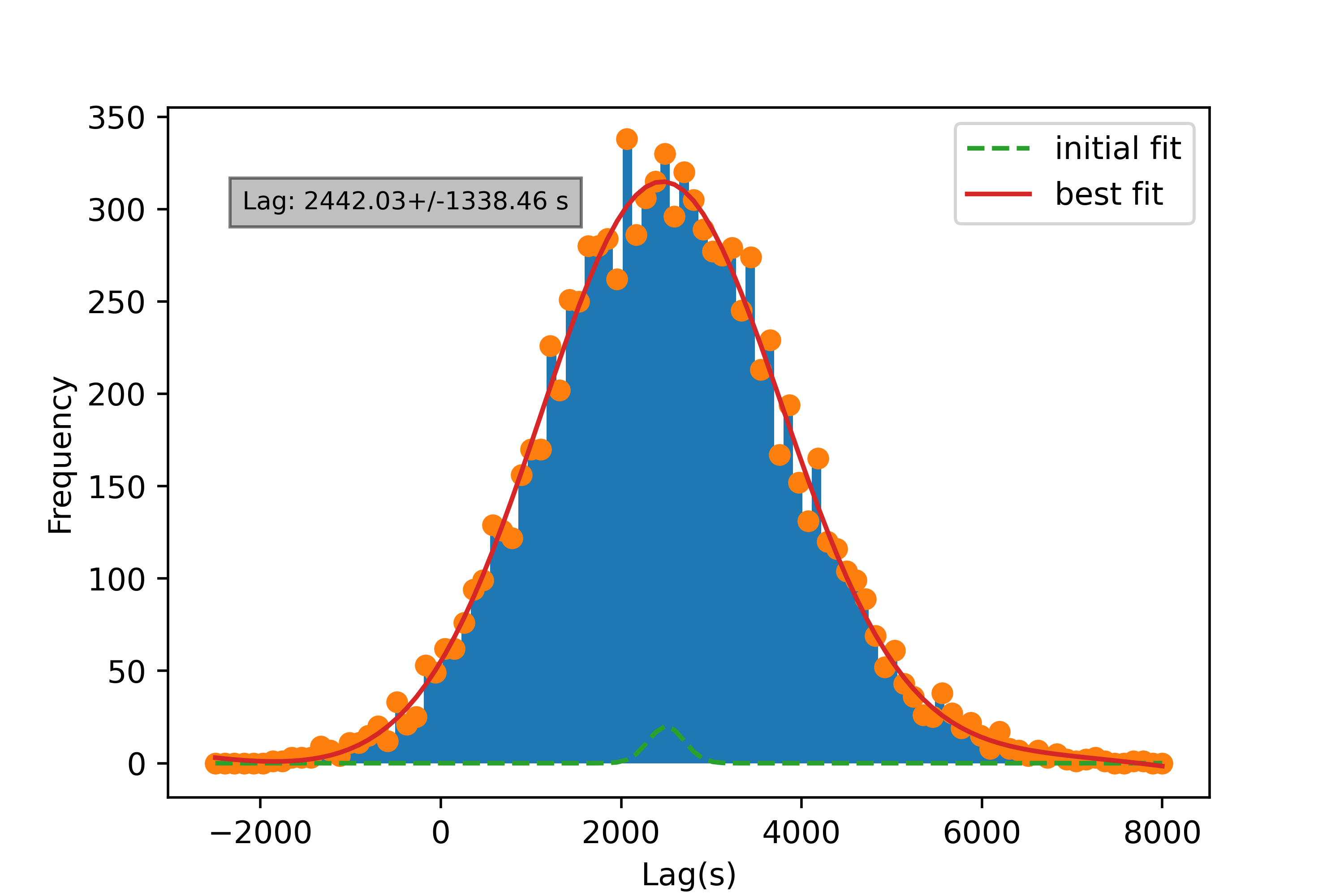}\\

\includegraphics[height=3 cm, width=8cm, angle=0]{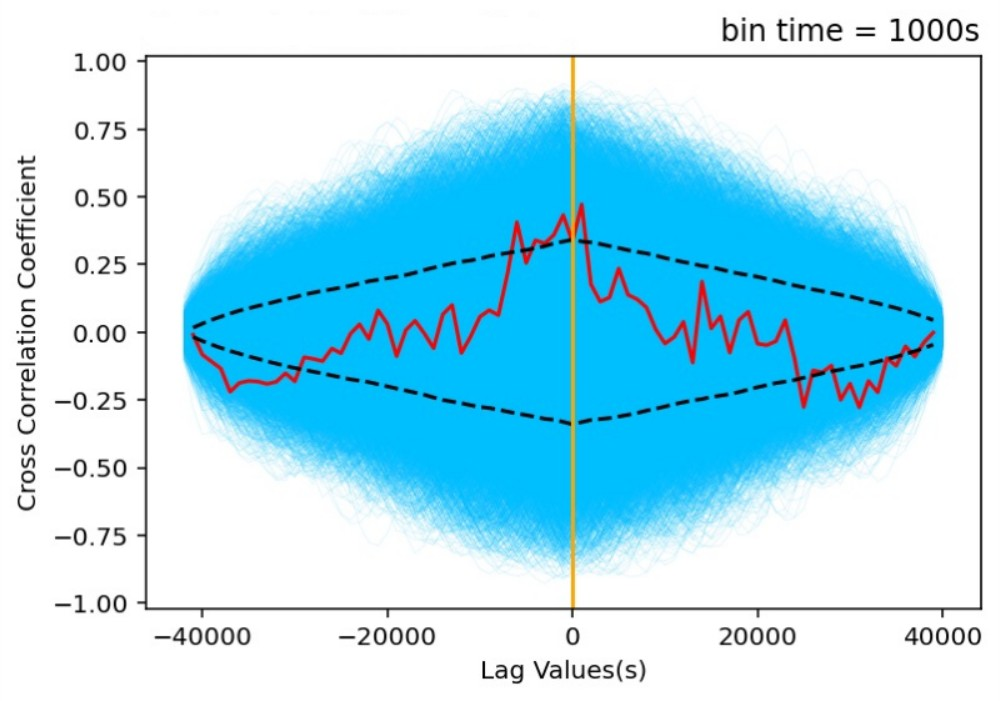} 
\includegraphics[height=3 cm, width=8cm, angle=0]{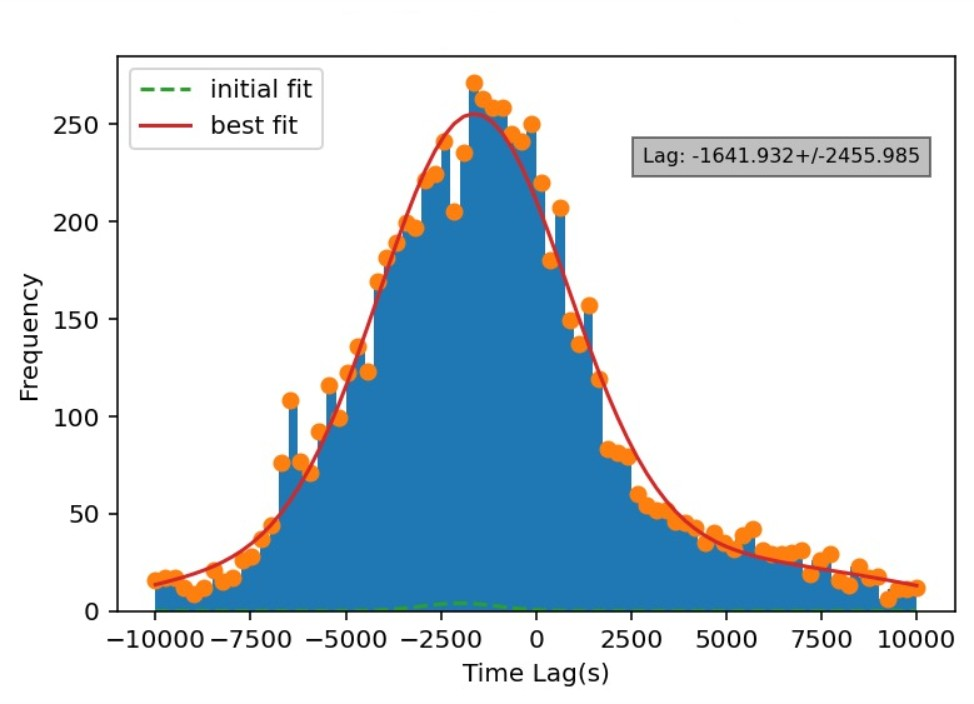}\\

\includegraphics[height=3 cm, width=8cm, angle=0]{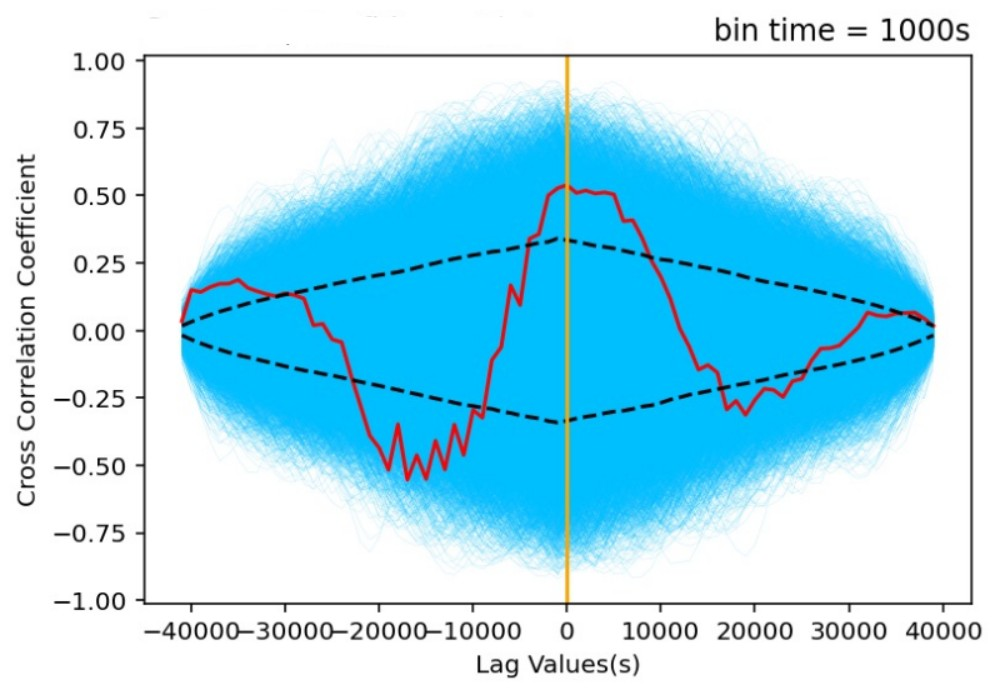} 
\includegraphics[height=3 cm, width=8cm, angle=0]{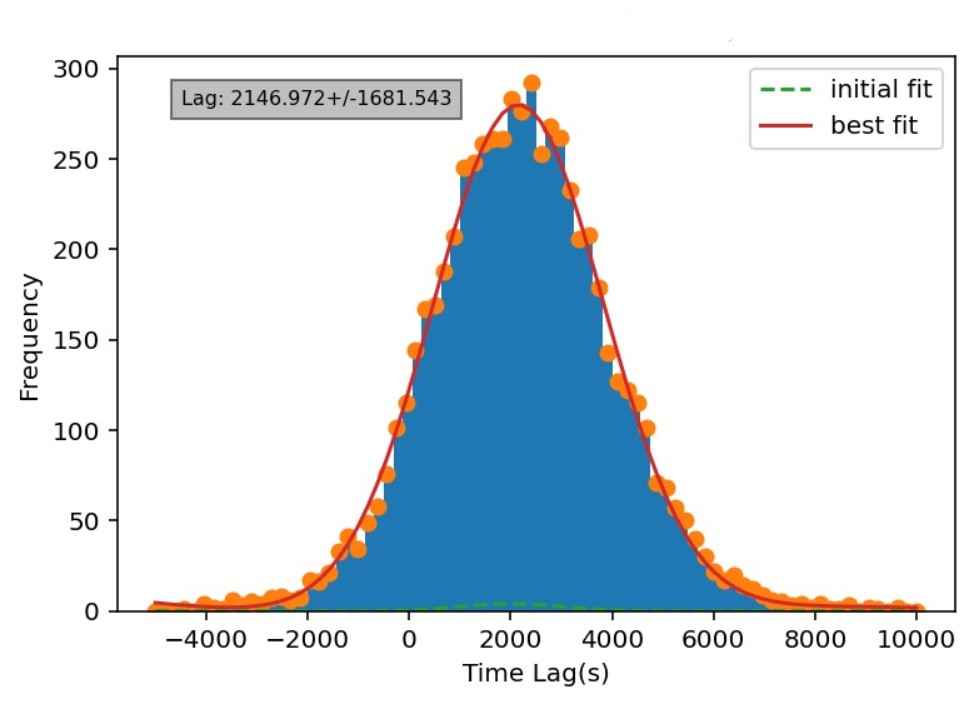}\\

\includegraphics[height=3 cm, width=8cm, angle=0]{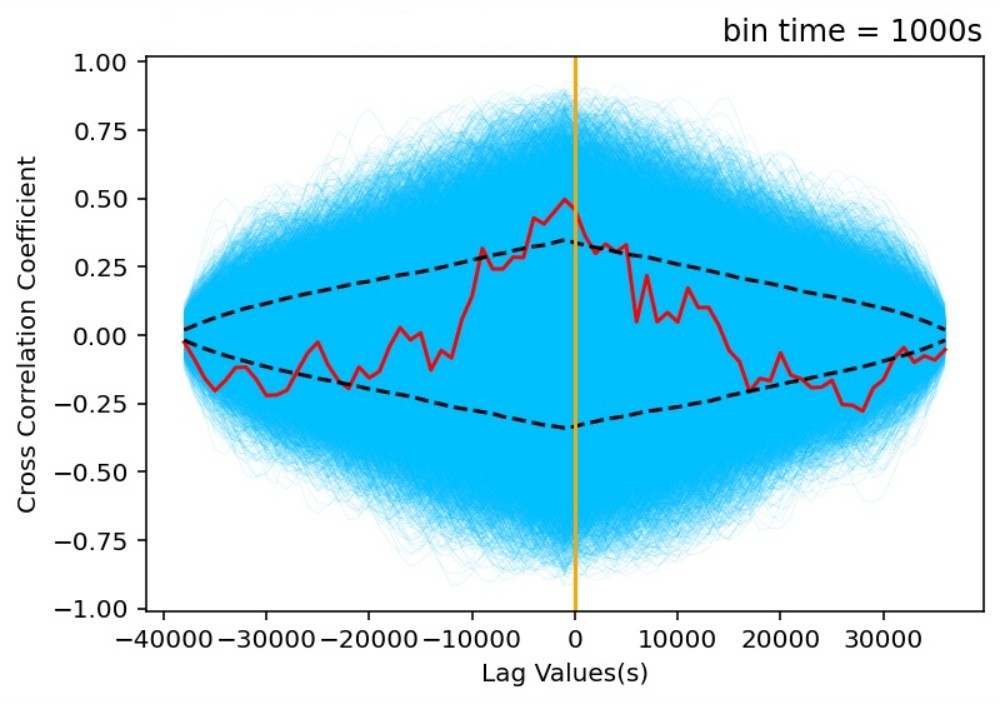} 
\includegraphics[height=3 cm, width=8cm, angle=0]{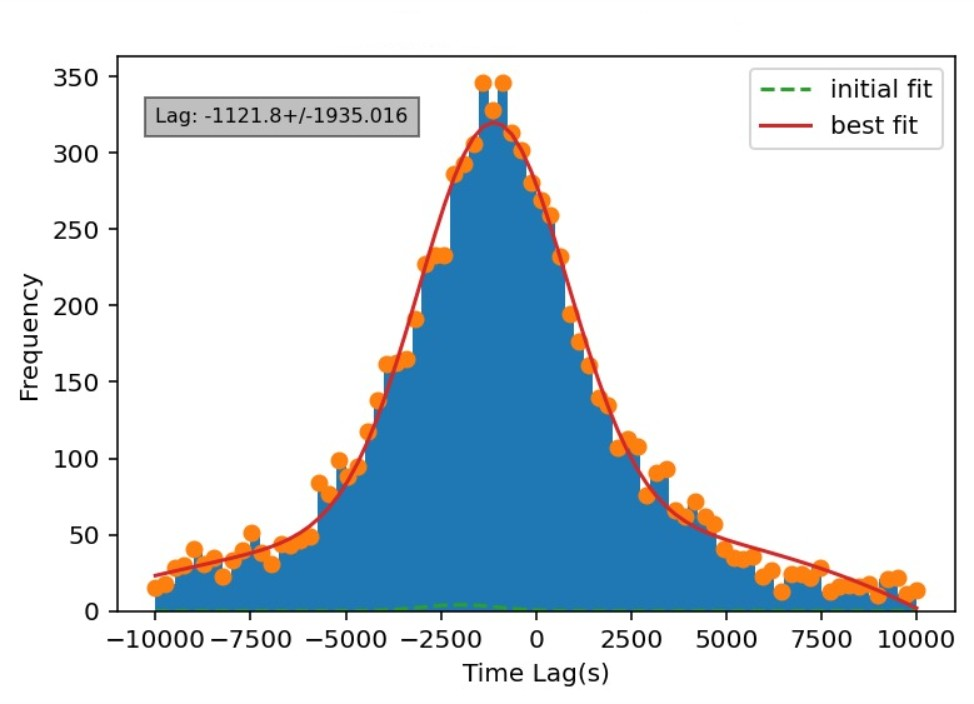}\\ 



\includegraphics[height=3cm, width=8.5cm, angle=0]{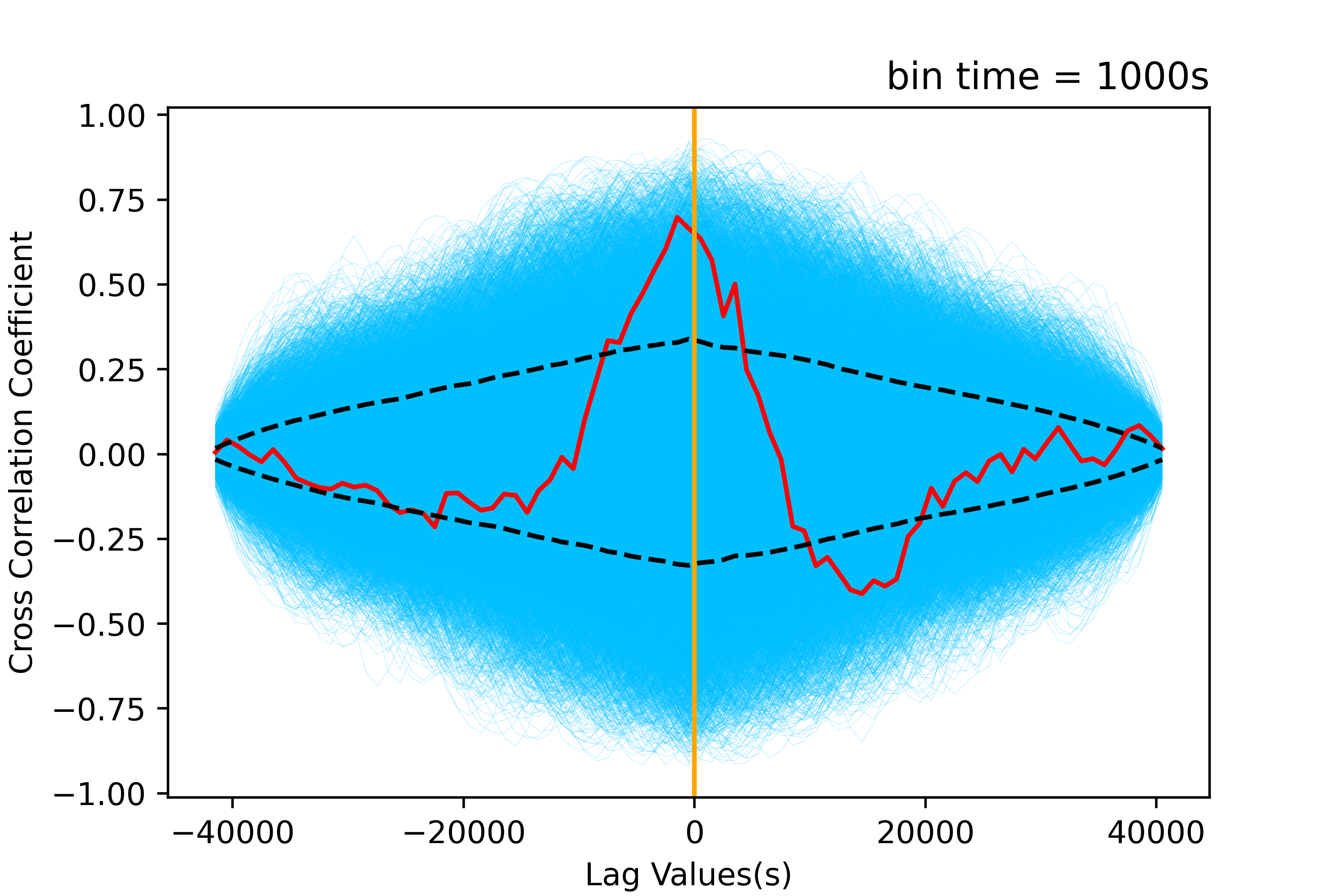} 
\includegraphics[height=3cm, width=8.2cm, angle=0]{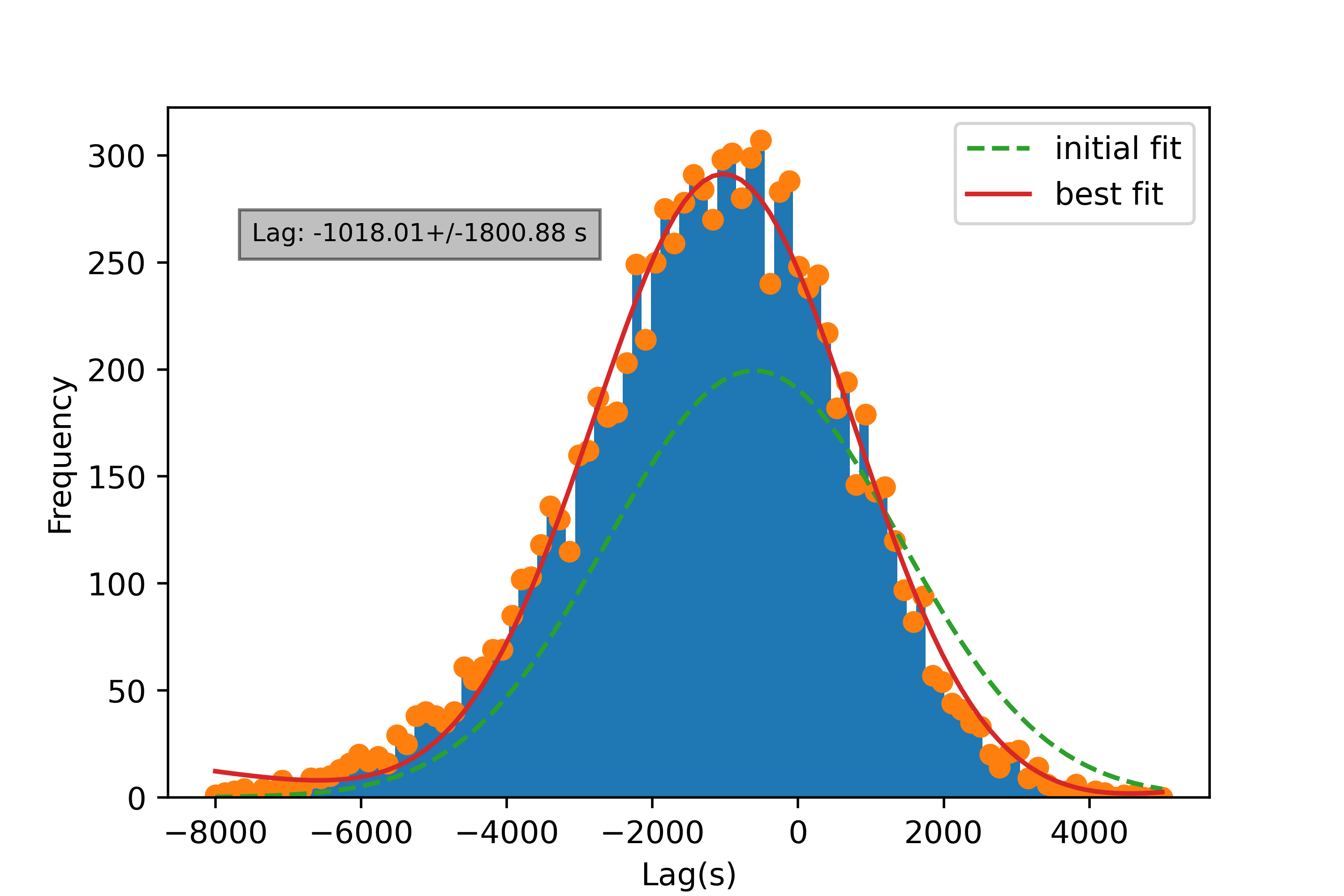}\\

\includegraphics[height=3cm, width=8.5cm, angle=0]{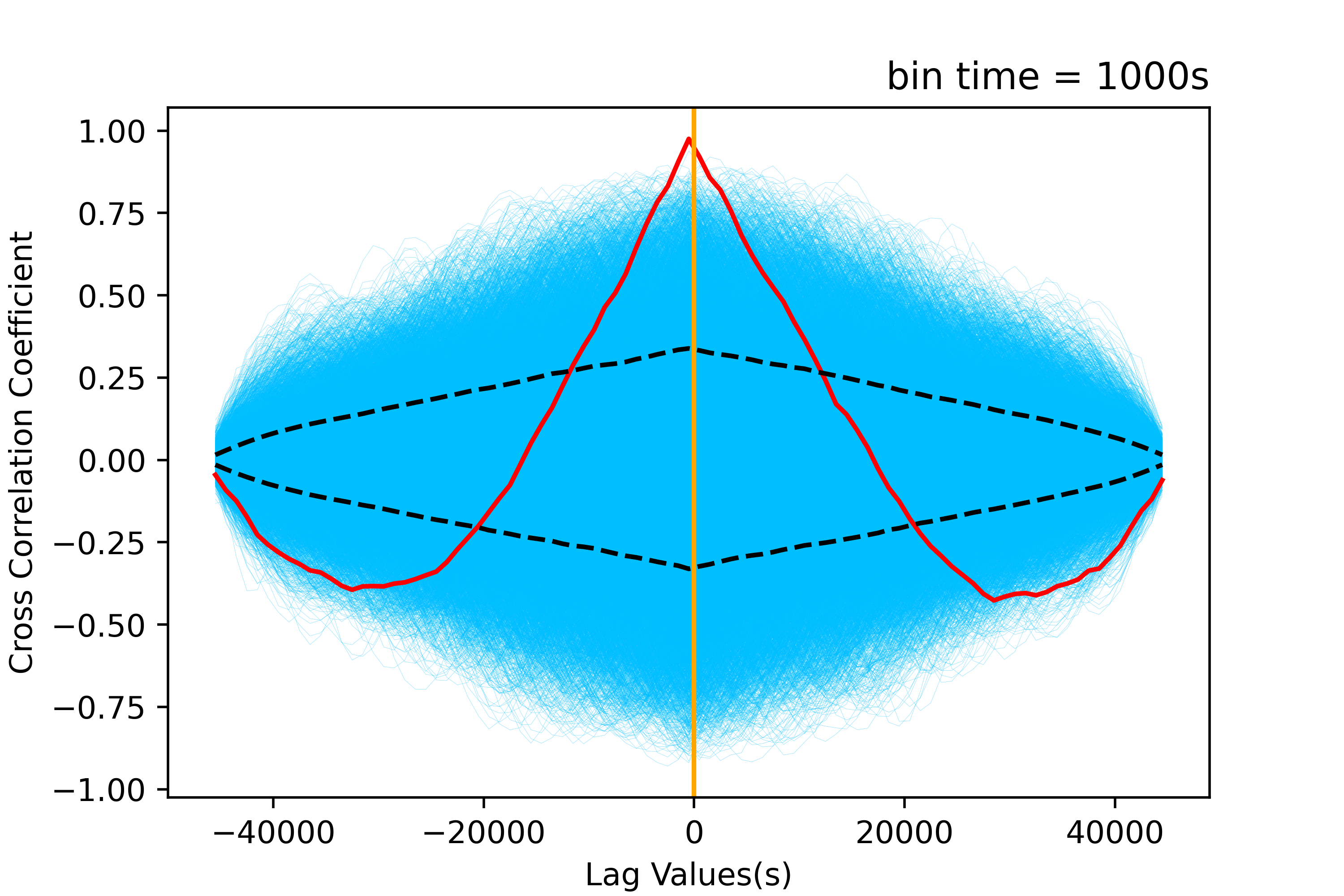} 
\includegraphics[height=3cm, width=8.2cm, angle=0]{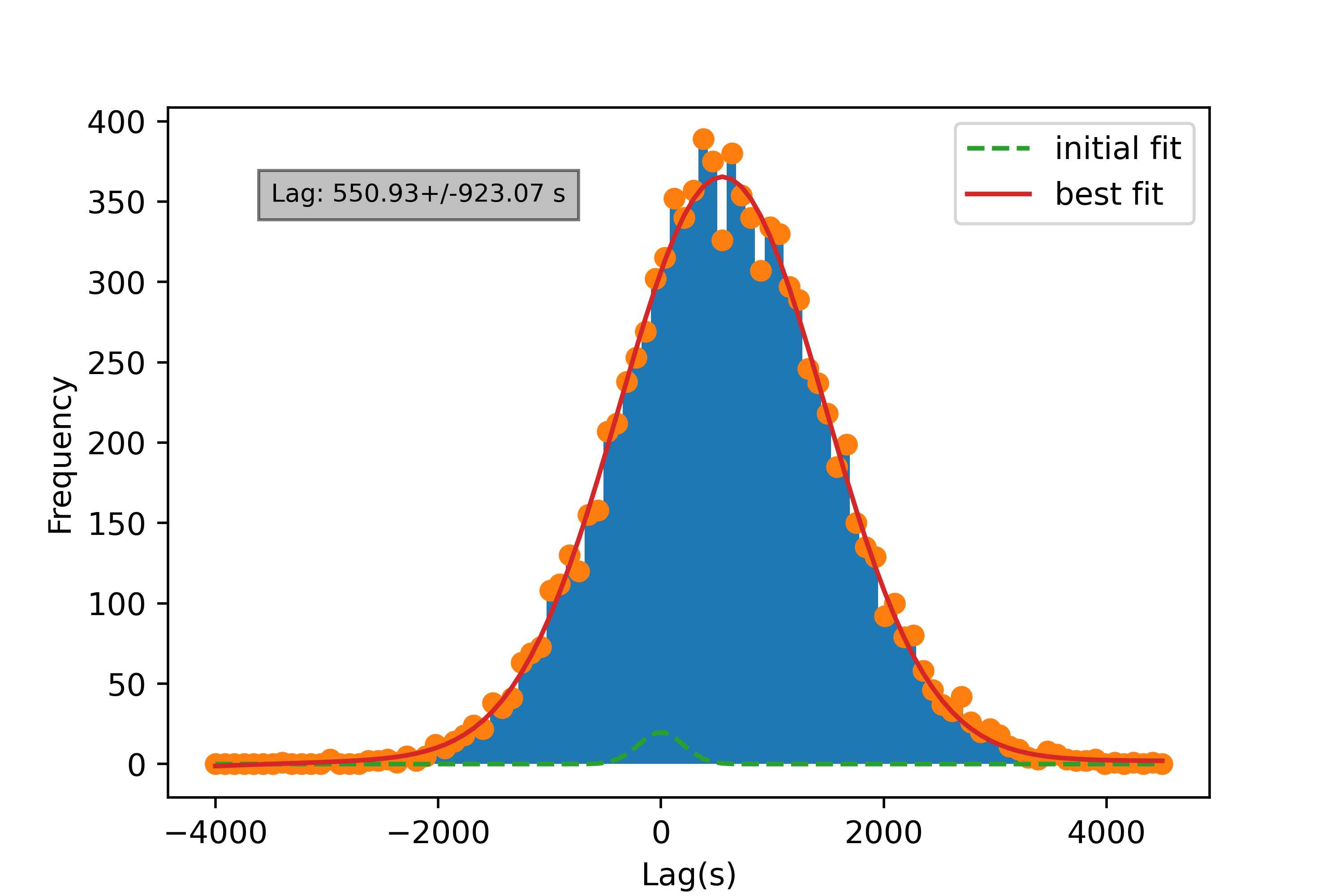}\\
\caption{Top panel: For Obs. 1 (2004), the blue shaded region shows the 10,000 simulated CCFs generated from the soft (0.3 -- 1 keV) and hard (1 -- 5 keV) light curves, with the 95\% significance levels indicated by dashed lines. The observed CCF is plotted in red, and the vertical dashed line marks zero lag. The corresponding lag histogram is shown in adjacent panels (see text for details). The green dashed line represents the initial fit parameters, which converge to the best-fit solution shown in red.
The subsequent panels present the CCFs and their lag histograms for Obs. 2a, 2b, 2c, and 2d (2019), followed by the results for Obs. 3 (2020). ///
\\Alt text:All the left panels consist of the confidence plot for the CCF and right side plots show the histogram plot }

\end{figure*}

\clearpage
\newpage

\begin{figure*}
\centering
\includegraphics[height=10cm, width=4cm, angle=270]{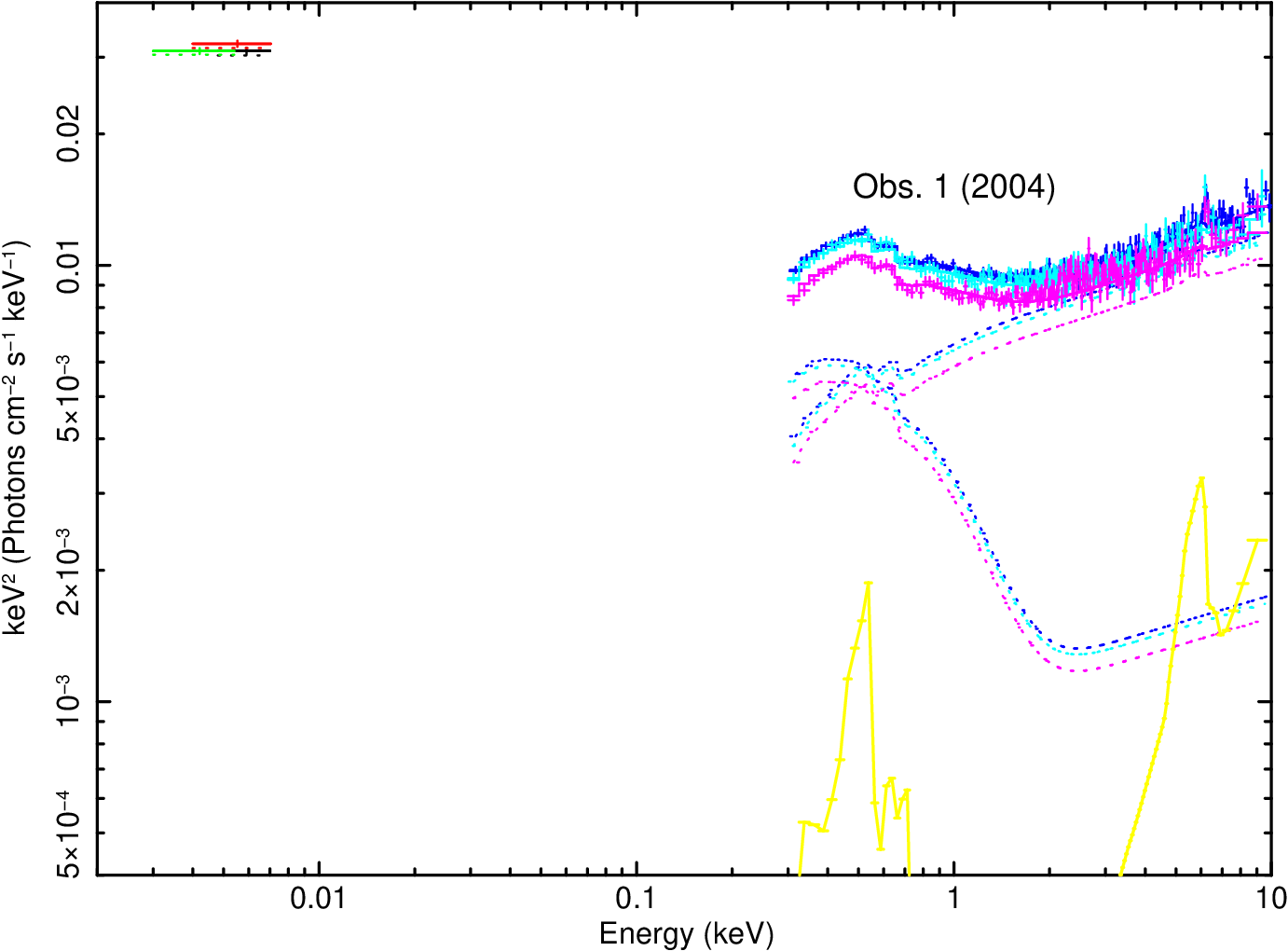}
\includegraphics[height=6cm, width=4cm, angle=270]{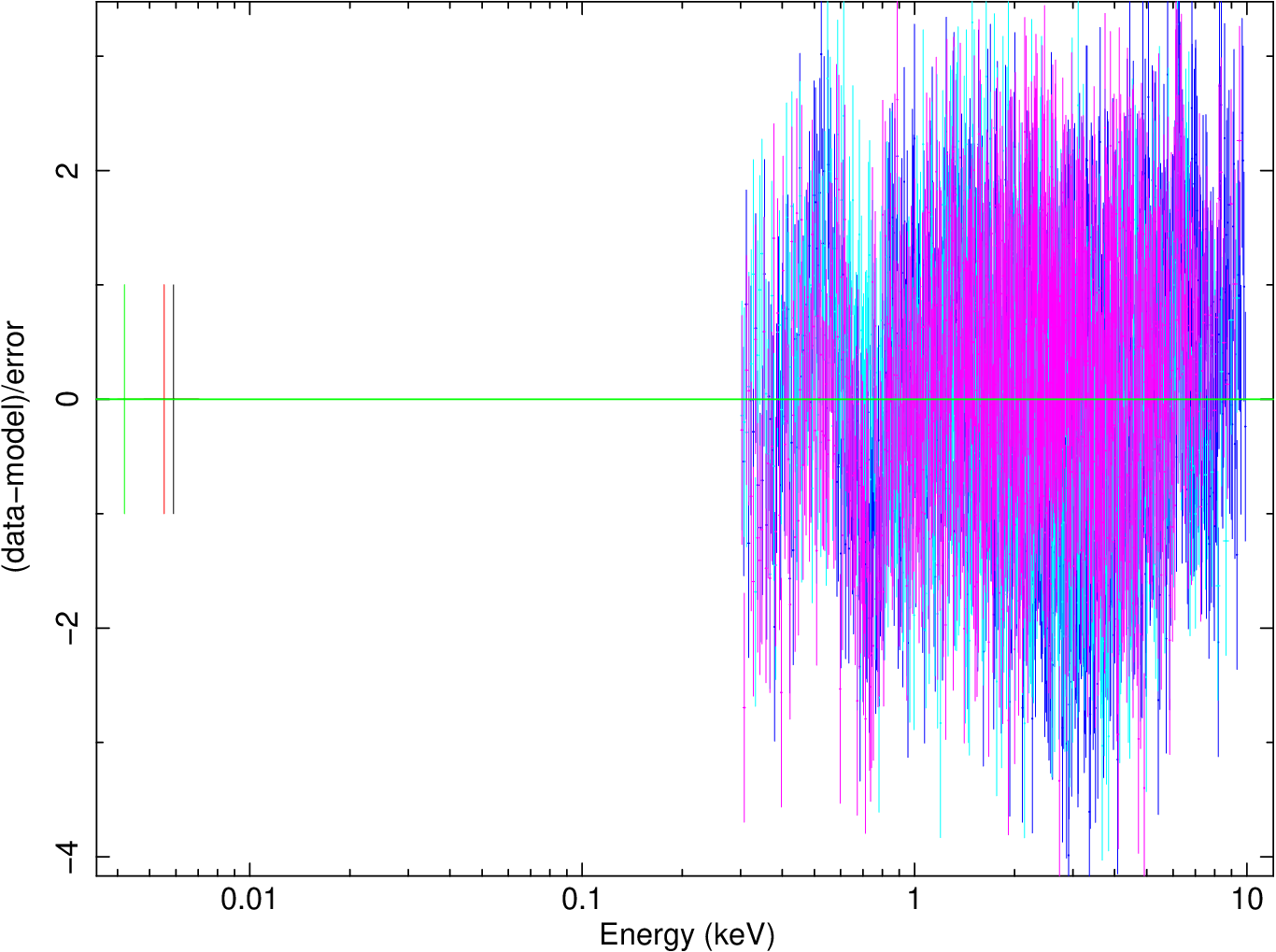}\\

\includegraphics[height=10cm, width=4cm, angle=270]{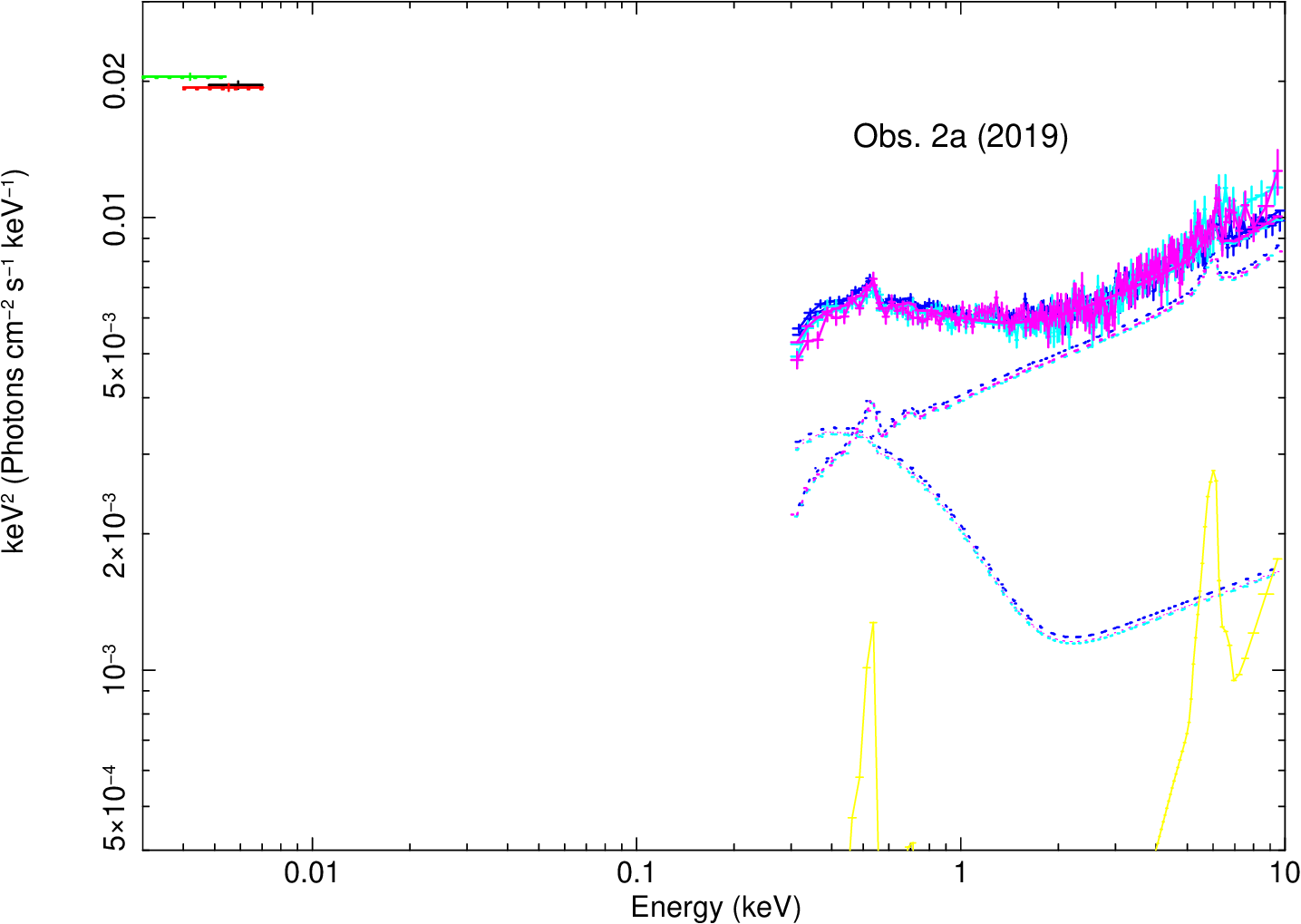}
\includegraphics[height=6cm, width=4cm, angle=270]{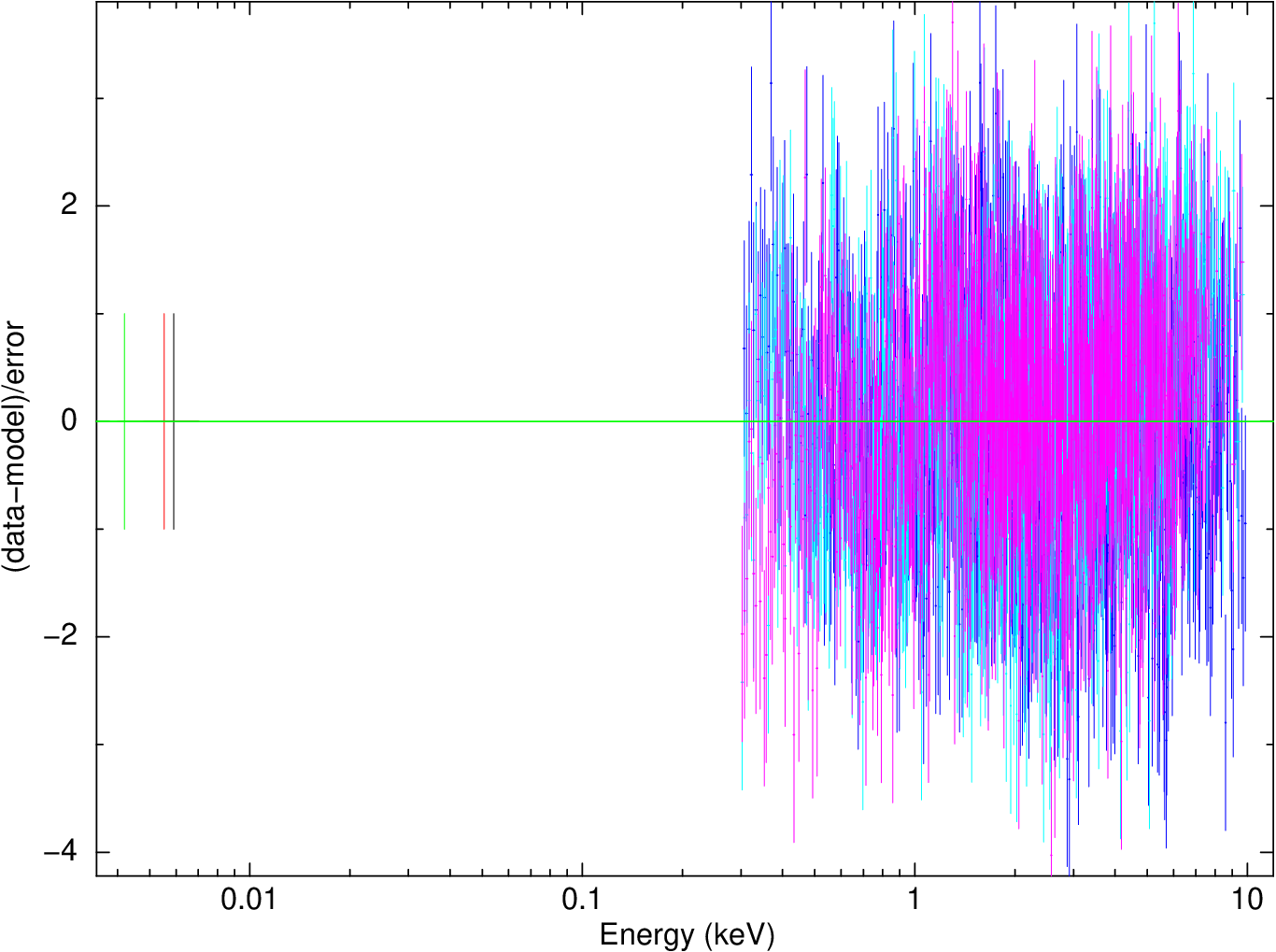}\\ 

\includegraphics[height=10cm, width=4cm, angle=270]{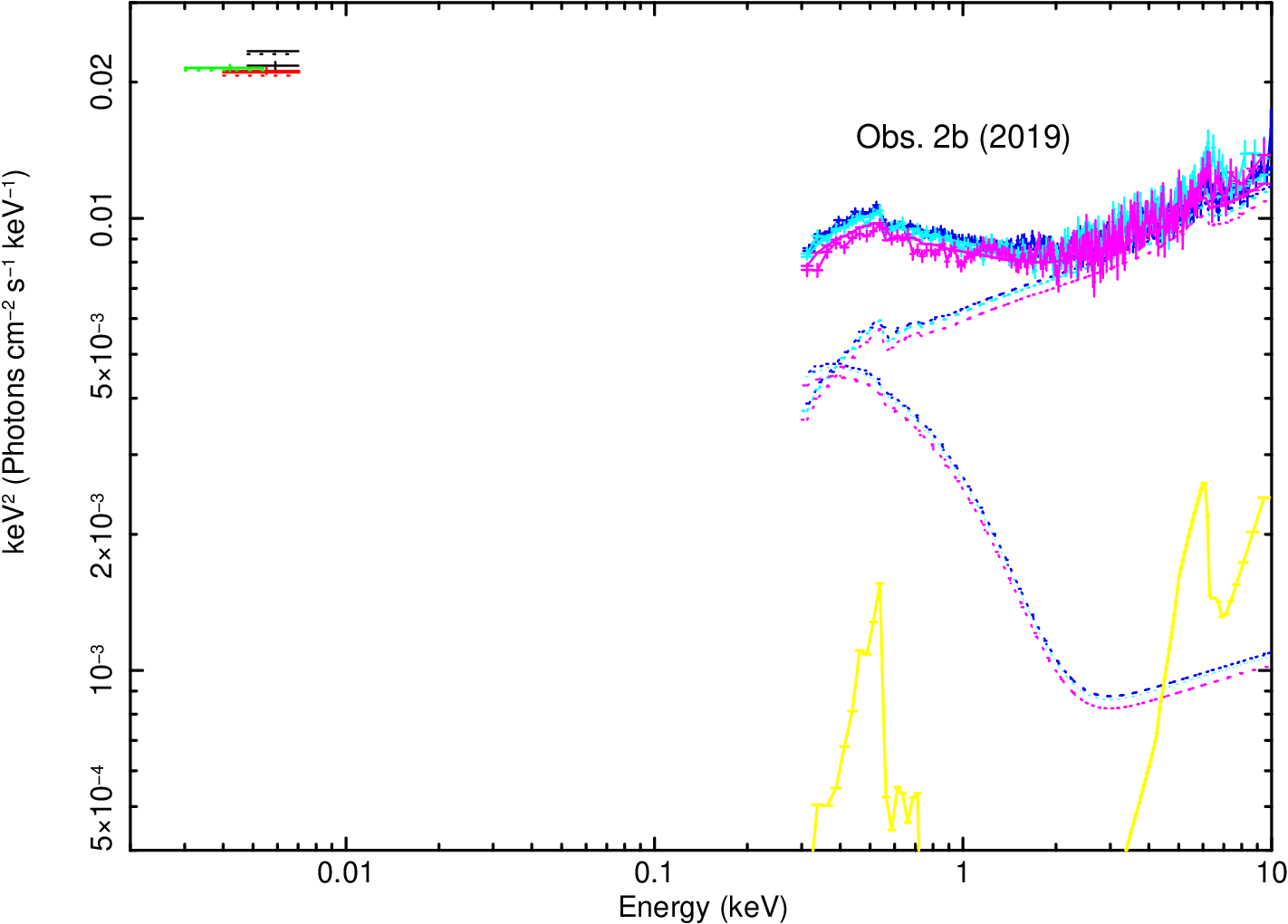} 
\includegraphics[height=6cm, width=4cm, angle=270]{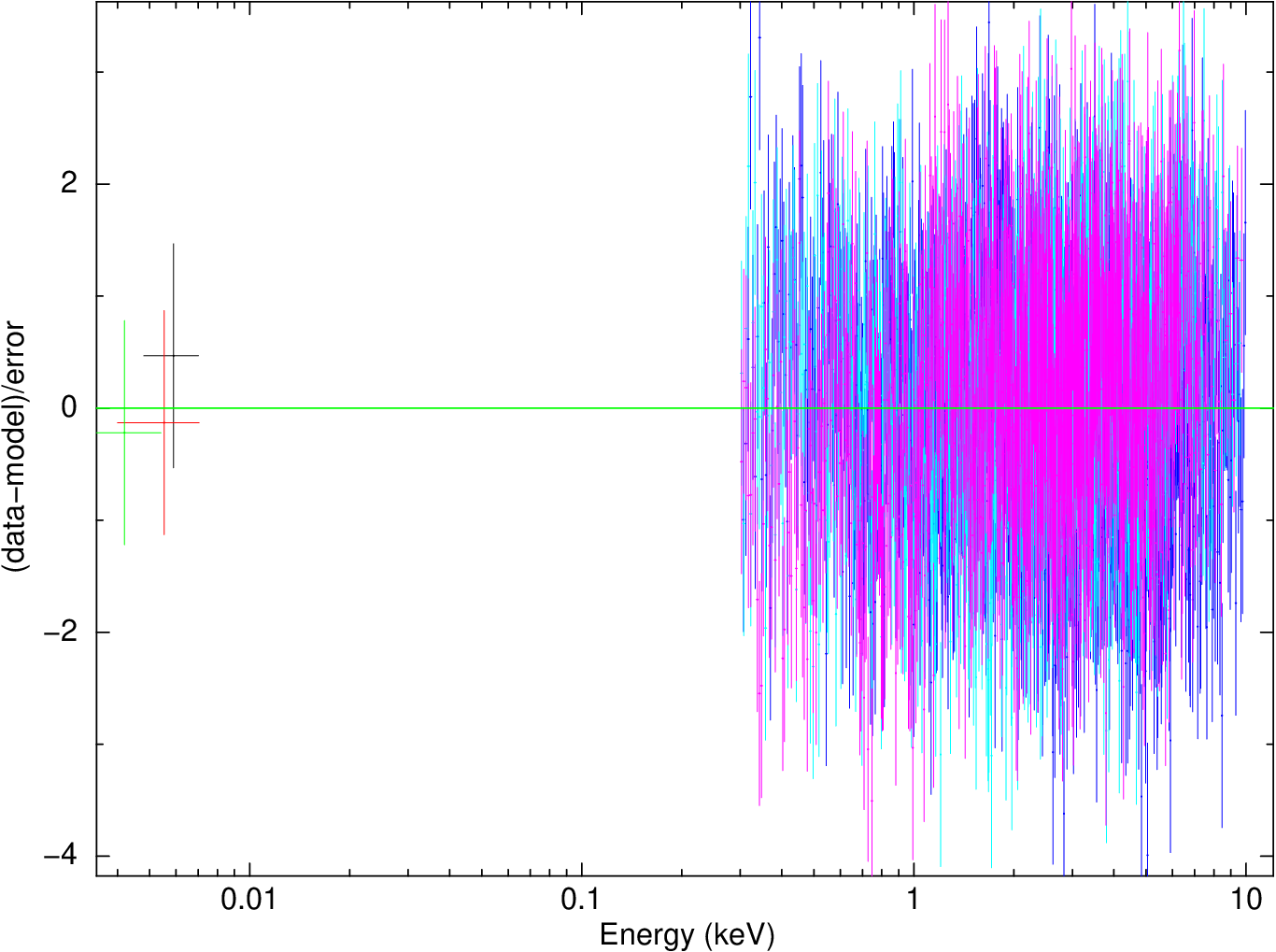}\\ 
\includegraphics[height=10cm, width=4cm, angle=270]{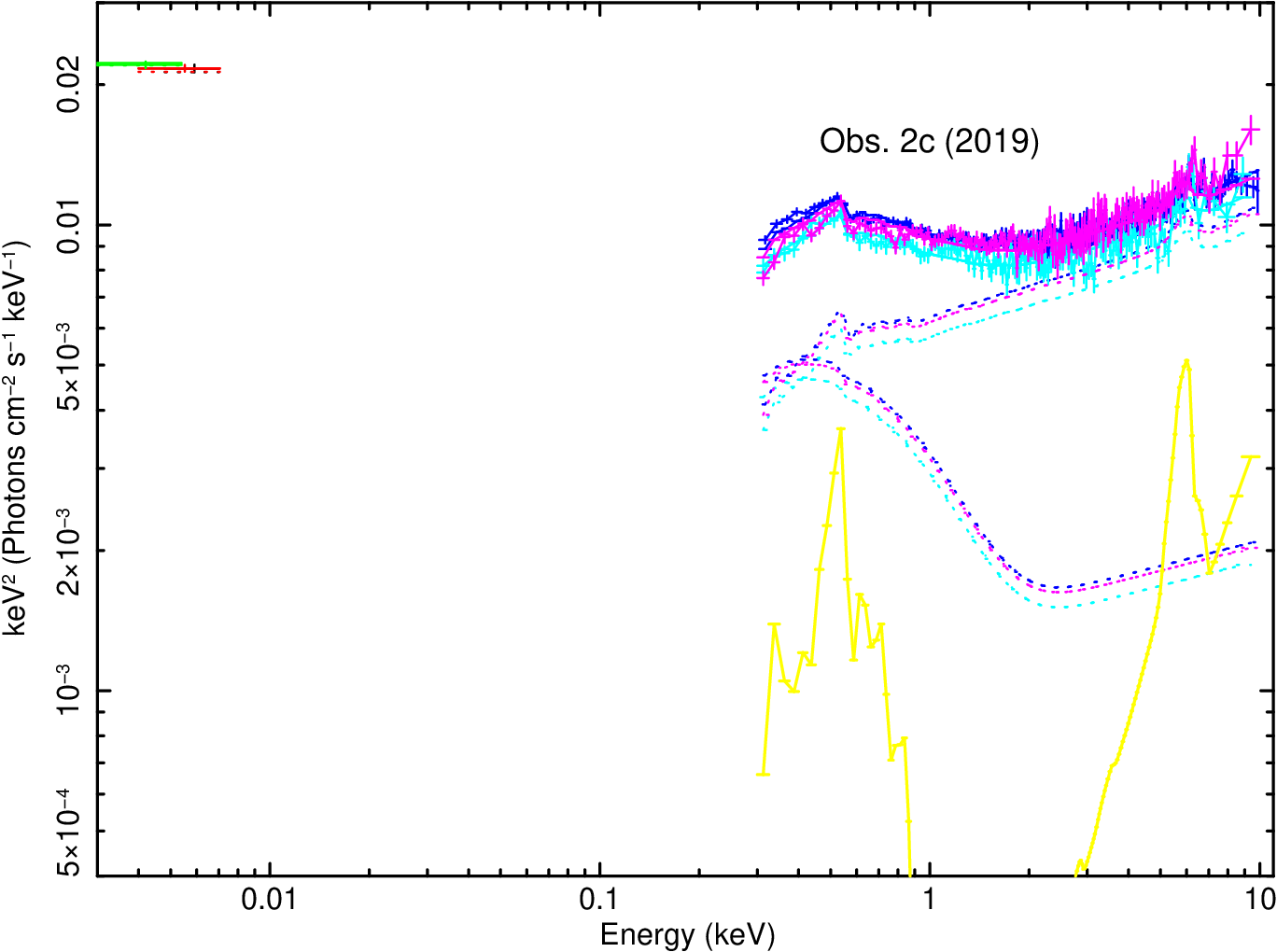} 
\includegraphics[height=6cm, width=4cm, angle=270]{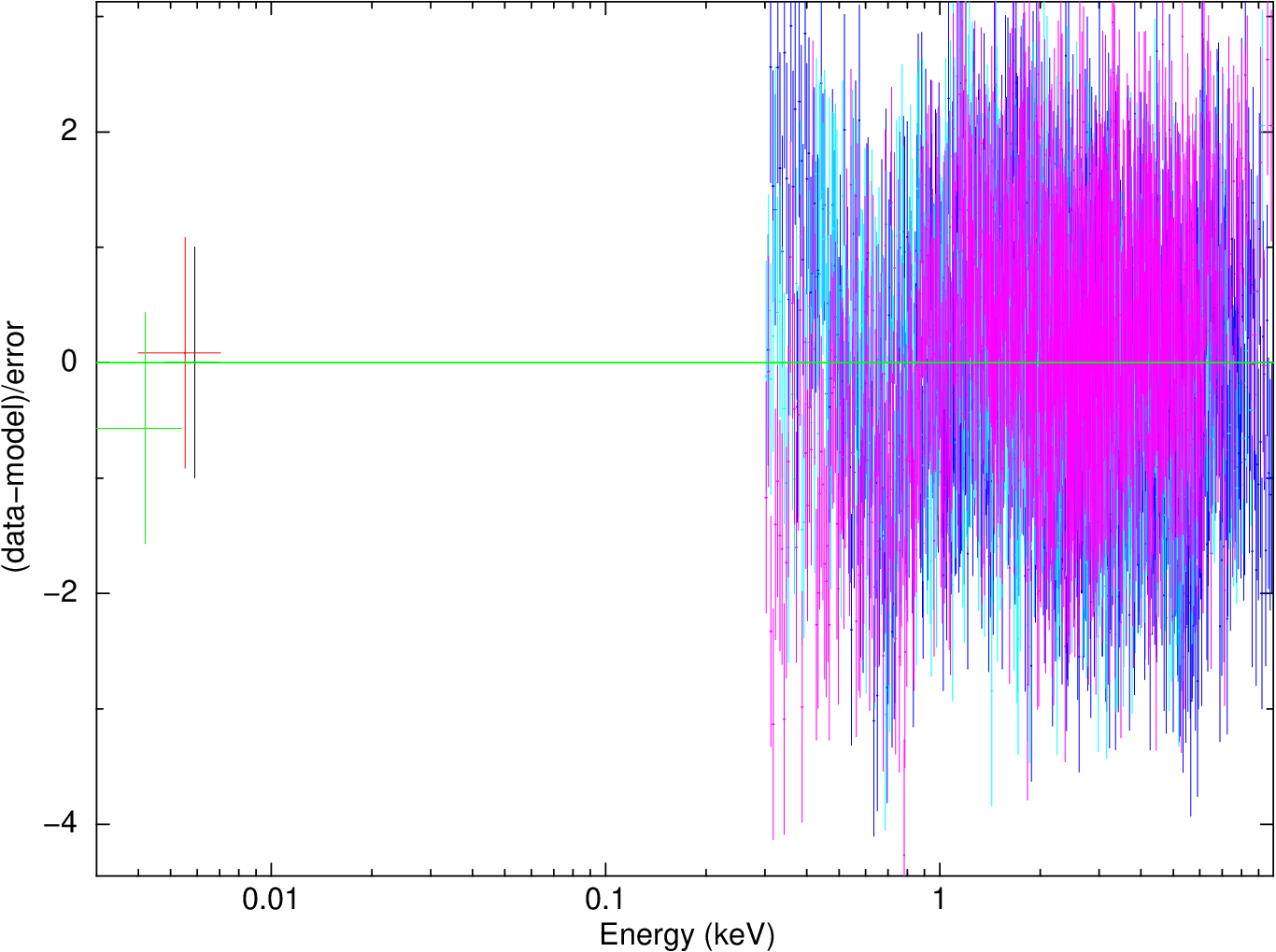}\\ 
\includegraphics[height=10cm, width=4cm, angle=270]{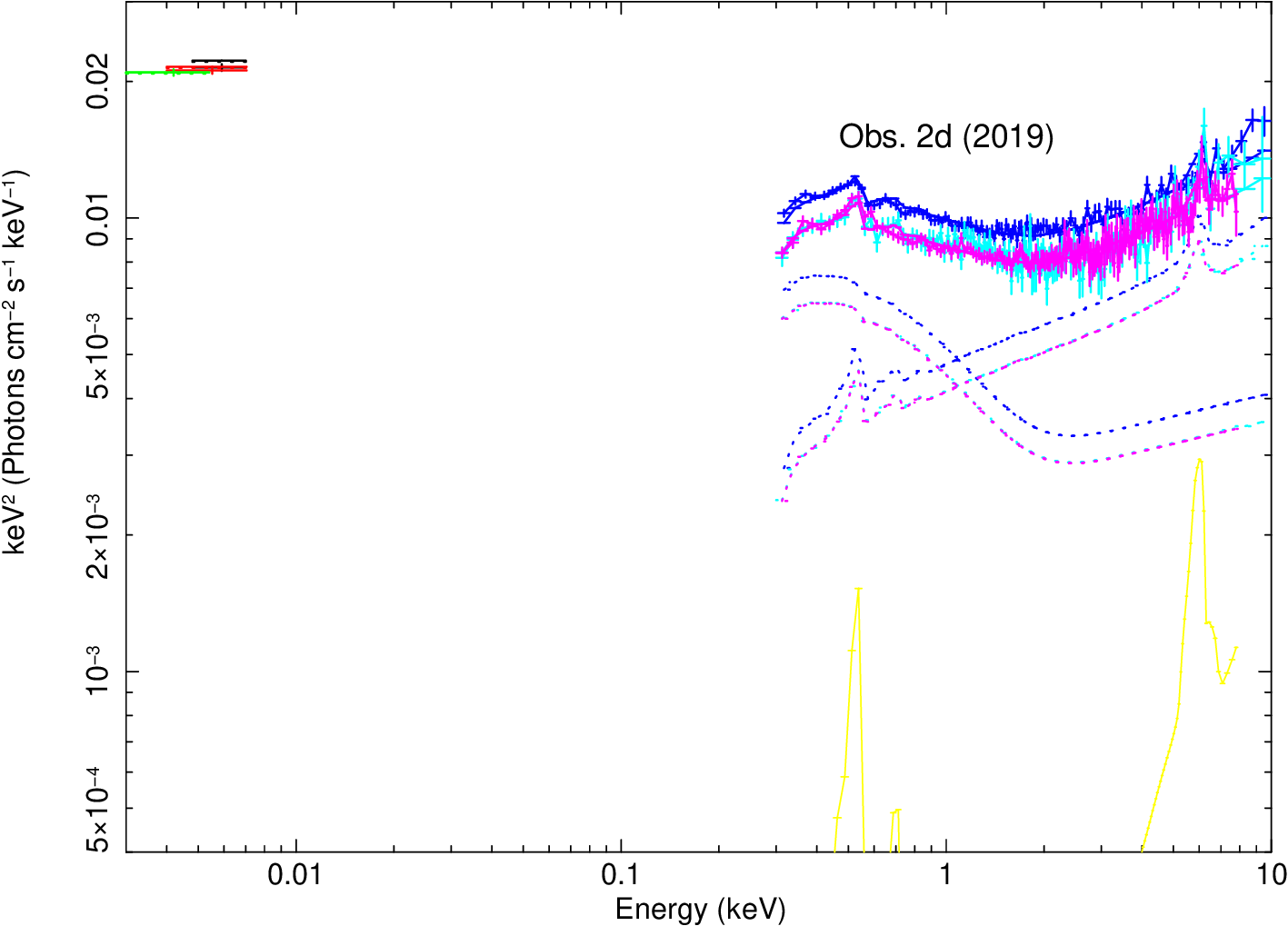}
\includegraphics[height=6cm, width=4cm, angle=270]{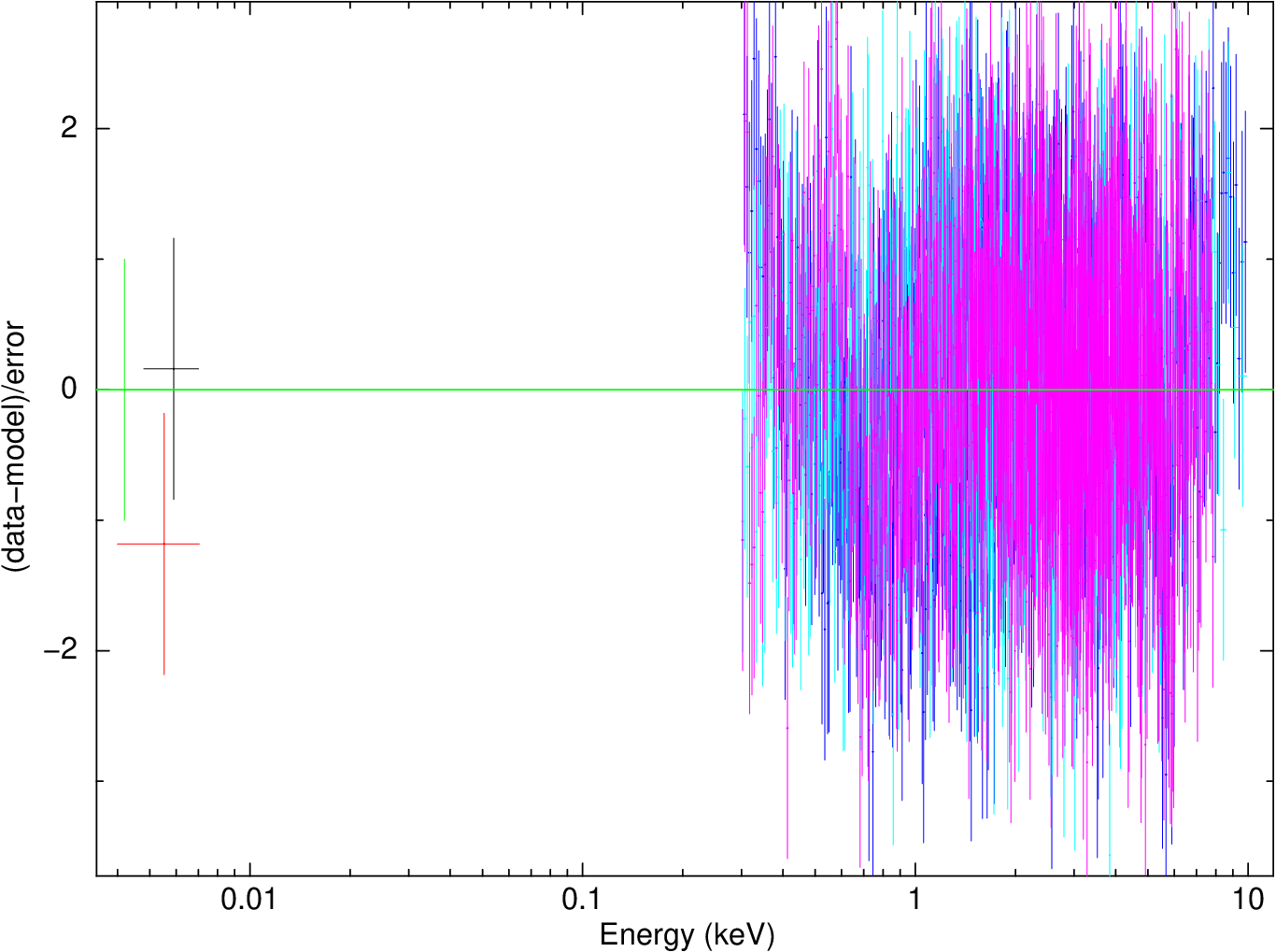}\\
\includegraphics[height=10cm, width=4cm, angle=270]{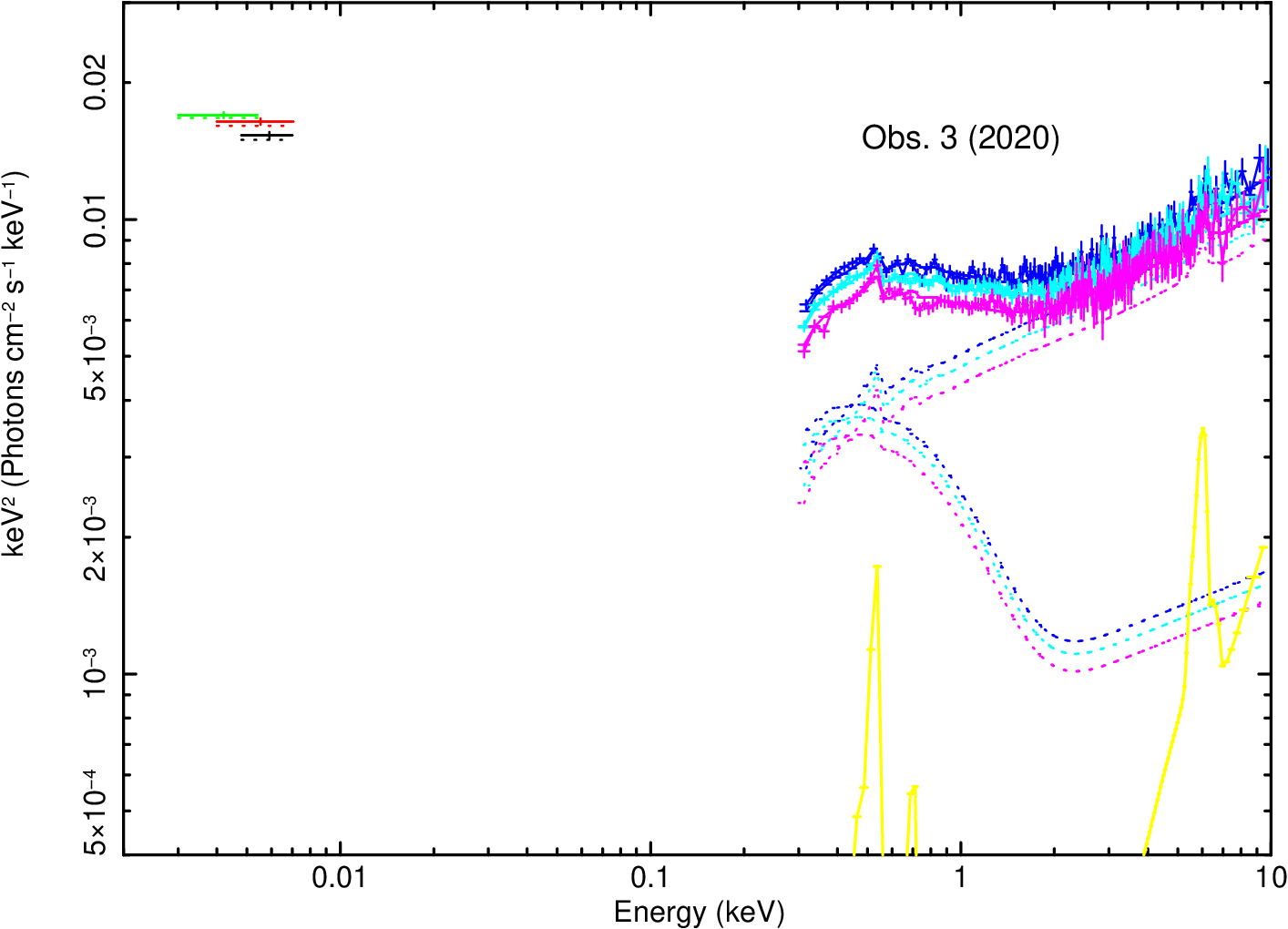}
\includegraphics[height=6cm, width=4cm, angle=270]{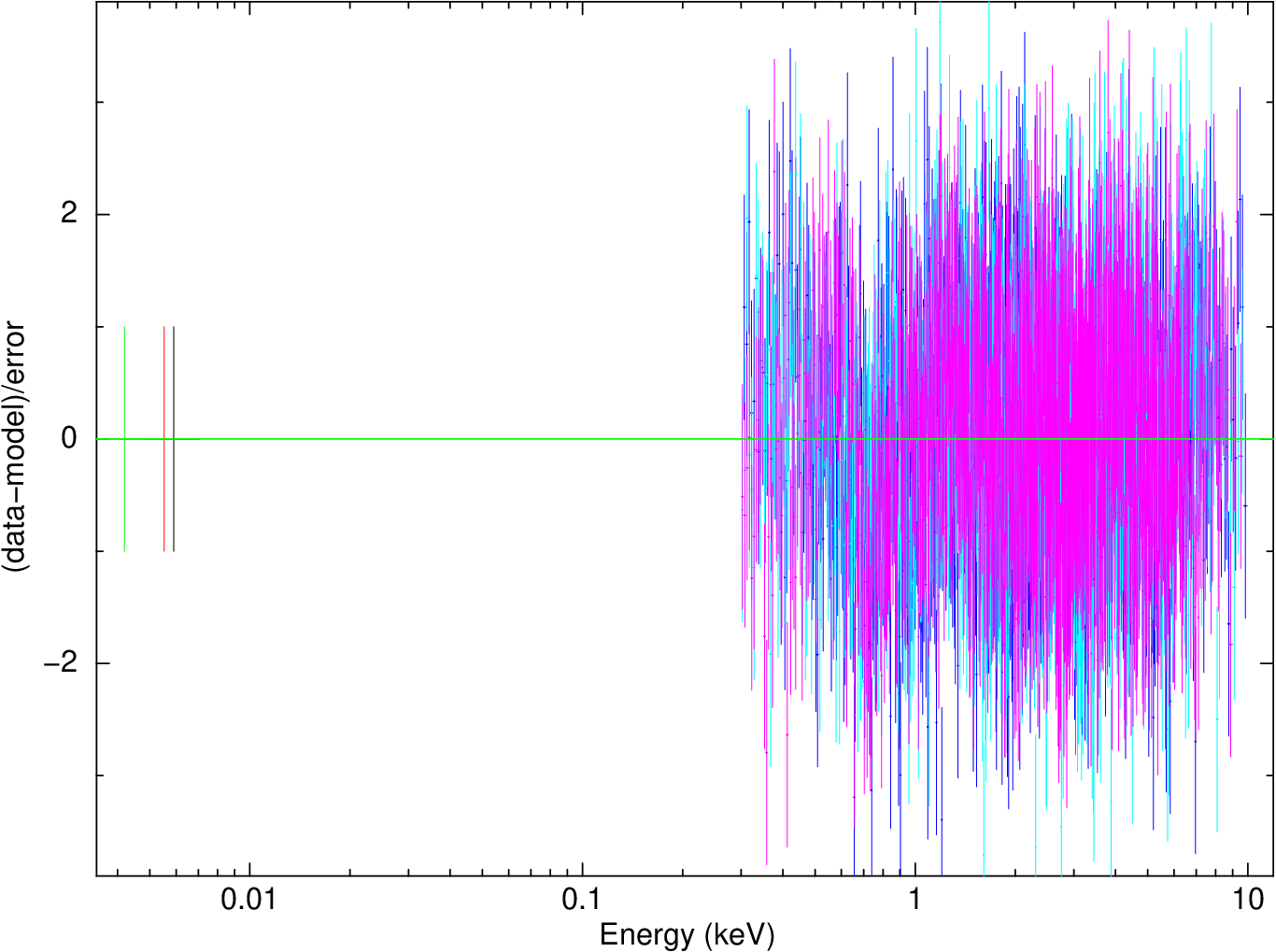}\\

\caption{Best fit spectra of all the observations using AGNSED+Relxillcp model (left panels, for M = 1.4 $\times$ 10$^{8}$ M$_{\odot}$) along with their respective residuals (right panels). The dotted dashed lines are the model components and yellow line displays the relativistic reflection component in each panel.///
\\Alt text: Energy spectra of all observations are presented here. }
    \label{fig:enter-label}
\end{figure*}

\section{Spectral Analysis}
{\bf To explore the broadband spectral properties of Mrk 110, we first applied the thermal Comptonization convolution model thcomp (Zdziarski et al. 2020), allowing us to characterize the X-ray power-law continuum and infer the physical properties of the corona. Accordingly, we adopted a two component Comptonization model i. e. 
\texttt{TBabs $\times$ Redden  $\times$ (thcomp(w) $\times$ zbbody + thcomp(h) $\times$ zbbody + ZGuass) },
where thcomp(w) describes the warm corona associated soft excess emission and thcomp(h) models the harder continuum. Since thcomp requires an extended energy grid, we expanded the model range to 10$^{-4}$ -- 10 keV using energies command in XSPEC. The Galactic absorption was fixed at N$_{\rm H}$ = 1.27 $\times$ 10$^{20}$ cm$^{-2}$, E(B-V) = 0.01 (Porquet et al. 2021) and the Gaussian centroid energy for Fe K$_{\alpha}$ line was fixed at E$_{c}$ = 6.4 keV.

The Comptonization parameters i. e. photon index ($\Gamma_{\rm h}$), electron temperature ($kT_{e}$), and covering fraction ($C_{\rm f}$) were left free during the fits. For the hot Comptonization component, the electron temperature was fixed at 30 keV (Porquet et al. 2024), and its seed photon temperature was tied to the warm corona temperature derived from thcomp(w). Covering fraction of thcomp(w) is kept at unity so that all the seed photons from the disk undergo Compton upscattering. Initially, the blackbody temperature kT$_{\rm bb}$ was left free and later fixed at 0.008 keV. The resulting best fit parameters are summarized in Table 1. The warm corona component (thcomp(w)) yields a temperature of $kT_{\rm w} \sim$ 0.33 keV with a steep spectral index of $\Gamma_{\rm w}$ $\sim$ 2.55 indicating towards a high optical depth $\tau_{\rm w}$ $\sim$ 15 except in Obs. 1 and 3 where $\Gamma_{\rm w}$ $\sim$ 2.35 was noted. The covering fraction of hot corona C$_{\rm f,h}$ was low during all the observations. We found that the blackbody temperature remained in the range kT$_{\rm bb}$ $\sim$ 7 -- 9 eV across all observations. Increasing this temperature beyond $\sim$ 0.2 keV leads to larger residuals in the spectral fits, indicating that higher values are disfavored by the data. 

Later to confirm the various parameters, we used another similar model, replacing the thcomp with CompTT components both for warm and hot corona and used diskbb model instead of zbbody i. e. \texttt{Tbabs $\times$ Redden $\times$ (diskbb+CompTT(w) + CompTT(h) + ZGauss)}. We noted inner disk temperature of kT$_{\rm in}$ $\sim$ 6 -- 8 eV in all the observations which was constrained after freezing a few other spectral parameters. The warm corona temperatures were 
kT$_{w}$ = 0.21$\pm$0.02 -- 0.24$\pm$0.01 keV with high optical depths of $\tau_{w}\sim$ 14 -- 15.  

{\bf Use of such a phenomenological model mentioned above is not purported to provide constraints on the locations of either corona} although the UV seed photons are assumed to originate in the outer accretion disk. A strong warm corona component can significantly modify the UV–X-ray continuum, motivating the use of a physically consistent broadband model. We therefore employed \texttt{Tbabs $\times$ Redden $\times$ (AGNSED + Relxillcp)}. In the AGNSED model (Kubota \& Done 2018), the accretion power released inside $R_{\rm h}$ generates the hard X-ray continuum through hot Comptonization. The region $R_{\rm h}$ $<$ R $<$$R_{\rm w}$ produces warm Comptonized emission, commonly associated with the soft excess, while the outer disk beyond $R_{\rm w}$ radiates as a standard thin disk. All three components are simultaneously computed, ensuring energy balance between them. To assess the presence of relativistic reflection, we incorporated Relxillcp (Dauser et al. 2014), in which the primary continuum is described by nthComp. The photon index and hot corona temperature were tied directly to the AGNSED values ($\Gamma_{\rm h}$, $kT_{\rm h}$). In Relxillcp model, the reflection fraction R$_{f}$ was kept free and is interpreted as the ratio of coronal photons illuminating the disk to those escaping to infinity. The ionization parameter (log $\zeta$ = 1), iron abundance (A$_{Fe}$ = 1), inclination ({\it i} = 10$^{\circ}$), and inner disk radius tied to R$_{hot}$ were fixed, while the emissivity profile was fixed to the canonical value q$_{1}$ = q$_{2}$ = 3 unless otherwise stated. We fixed the black hole mass and spin to $M_{\rm BH}$ = 1.4 $\times$ 10$^{8}$ M$_{\odot}$
and a$_{spin}$ = 0.84 (Porquet et al. 2024). The remaining fixed parameters are summarized in Table 2. For each observation, we allowed the mass accretion rate ${\dot{m}}$, the hot and warm corona electron temperatures ($kT_{\rm h}$, $kT_{\rm w}$), their corresponding spectral indices ($\Gamma_{\rm h}$, $\Gamma_{\rm w}$), and the outer radii of both Comptonizing regions ($R_{\rm h}$, $R_{\rm w}$) to vary along with the reflection fraction (R$_{f}$) of Relxillcp. All best fit values and uncertainties at 90\% are listed in Table 2, and spectra for all the observations are shown in Fig. 8.
We find that the mass accretion rate is highest in Obs. 1 and Obs. 2d (log ${\dot{m}}$ $\sim$ $-$1.48) and lowest in Obs. 2a (log 
${\dot{m}}$ $\sim$ $-$1.60). The hot corona photon index remains approximately constant at 
$\Gamma_{\rm h}$$\sim$ 1.83 except for Obs. 2a and Obs. 3 which were at $\Gamma_{\rm h}$$\sim$ 1.73, while the warm corona index varies between 
$\Gamma_{\rm w} \sim$ 2.14 -- 2.32. The size of the warm corona is generally constrained to $R_{\rm w}$ $\sim$ 100 $R_{\rm g}$, except in the highest luminosity epoch (Obs. 2d), where it increases to $R_{\rm w}$ $\sim$ 183 $\pm$ 38 $R_{\rm g}$. The hot corona remains compact across Obs. 2a--2d 
($R_{\rm h} \sim 7-15 R_{\rm g}$). 

Importantly, the reflection fraction remains consistently low, R$_{f}$ $\sim$ 0.2 -- 0.8, across all epochs. The modest differences between our derived coronal sizes and those of Porquet et al. (2024) are likely due to the absence of NuSTAR coverage in our fits. Nevertheless, the relative trends in $R_{\rm h}$, $R_{\rm w}$, and ${\dot{m}}$ are in good agreement, reinforcing the robustness of our results.

We also repeated the entire analysis assuming a lower black hole mass
($M_{\rm BH}$ = 2 $\times$ 10$^{7}$ M$_{\odot}$). The fits show systematically higher accretion rates (log ${\dot{m}}$ $\sim$ $-$0.43 to $-$0.70; Table 2, bottom), slightly higher warm corona temperatures (kT $_{\rm w}$ $\sim$ 0.33 keV), a slight variation of hot and warm coronal indices were noted i. e. ($\Gamma_{\rm h}$ $\sim$ 1.70 $\pm$ 0.02 -- 1.84 $\pm$ 0.02, $\Gamma_{\rm h}$ $\sim$ 2.41$\pm$0.05 -- 2.61 $\pm$ 0.04). The coronal sizes were around $R_{\rm h}$ $\sim$ 10–20 $R_{\rm g}$, $R_{\rm w}$ $\sim$ 160 -- 213 $R_{\rm g}$, but again the reflection fraction remained low (R$_{\rm f}$ $<$ 1), consistent with that of high mass BH spectral fits.

To examine potential degeneracy with the emissivity index, we allowed q$_{1}$ to vary freely. It steepened to values 5 -- 8 across the observations, yet the reflection fraction remained low (R$_{\rm f}$ $<$ 1) and emissivity indices could not be constrained properly. 
To confirm the robustness of the observed low reflection fraction, we varied q$_{\rm 1}$ from 3 -- 8, allowing the R$_{f}$ to vary. In Obs. 2d, for q$_{\rm 1}$ = 3 and R$_{\rm f}$ = 0.8, we obtained  $\chi^{2}$/dof = 0.93 while a steeper emissivity q$_{1}$ = 8 resulted in a similar $\chi^{2}$/dof = 0.96 with R$_{\rm f}$ = 0.5. But when the reflection fraction was increased to 
R$_{\rm f}$ = 2 or above, the fit degraded, resulting in $\chi^{2}$/dof = 1.45 for q$_{1}$ = 3 and $\chi^{2}$/dof = 1.49 for q$_{\rm 1}$ = 8. The F-test statistics for q$_{1}$ = 3 and R$_{f}$ from 0.5 to 2 resulted in F$_{\rm stat}$ = 3009 with a probability 2 $\times$ 10$^{-236}$. Similar variation was observed in the remaining datasets. Overall it indicates that steepening the emissivity profile (from q$_{ \rm 1}$ = 3 to 8) does not enhance the contribution of the reflection component, and it remains to be low R$_{\rm f}$ $<$ 1.}

\section{Results and Discussion}
Mrk 110 is a bare AGN where intrinsic absorption is absent hence, studying such a system will provide a clear picture of the underlying physical and radiative mechanisms in the inner region of the accretion disk. We noted a soft lag of the order of a few hundred to thousand of seconds (see section 3.2) and the predicted soft lag is $\tau$ = 282 -- 730 s for a M = 1.4$\pm$0.3$\times$10$^{8}$ M$_{\odot}$ which should occur at $\nu$ = 6.1$\times$10$^{-5}$ -- 1.4$\times$10$^{-4}$ Hz (De Marco et al. 2013). {\bf However, we detect a promising soft lag of $\tau$ = $-$ 3067 $\pm$ 1412 s, which appears to deviate from the expectation based on the empirical mass vs soft lag relation reported by De Marco et al. (2013). For Mrk 110, considering a black hole mass of 
M = 2$\pm$0.3$\times$10$^{7}$ M$_{\odot}$, De Marco et al. (2013) report much longer soft lag of $\tau$ $\le$ $-$5000 s with a significance of $<$1 $\sigma$  (see their Fig. 6), indicating a clear deviation from the simple scaling, which predict  soft lags of only $\sim$ 111 -- 186 s.} 
The CCF shows soft lag in Obs. 2 (a, c, d) viz. ($\tau_{ccf}$ = $-$1642 -- $-$1081 s; Fig. 7, middle panels) and hard lag was detected in other observations which is consistent with the results reported by Dasgupta \& Rao (2006) for Obs. 1. Nevertheless, few observations does provides evidence for a soft lag in both the frequency–lag and CCF analyses for Mrk 110. The energy–lag spectra display reverberation like trend, that appears in the same frequency range ($\sim$ 9$\times$10$^{-5}$ Hz) where our frequency–lag analysis reveals a soft lag (Figure 6; Obs. 1 and 2). 
Hence the frequency and energy lag spectra supports the reverberation scenario in Mrk 110.



The soft lags in AGNs are frequently interpreted as evidence for relativistic reflection from the innermost disk (Fabian et al. 2009; Kara et al. 2013; Kara et al. 2014). 
{\bf A soft lag of $\tau$ = $-$ 3067 $\pm$ 1412 s corresponds to a light travel distance of approximately R $\sim$ 4.5$^{+2.0}_{-2.1}$ $R_{\rm g}$ for a black hole mass of M = 1.4$\pm$0.3$\times$10$^{8}$ M$_{\odot}$, placing the inferred reverberation radius close to the ISCO. Consequently, a pure reflection driven reverberation origin for the observed soft lags is entirely viable. But when M = 2$\pm$0.5 $\times$ 10$^{7}$ M$_{\odot}$, the inferred light-travel distance of R $\sim$ 31$^{+15.0}_{-15.0}$ $R_{\rm g}$ indicates that the observed lag is likely influenced by a combination of relativistic reverberation and variability associated with the warm corona, rather than being produced by a single mechanism alone. However, the limited signal to noise warrants cautious interpretation.}

{\bf 
 Although, the present study of Mrk 110 does not reveal spectral signatures indicative of a strongly reflection dominated geometry. The Fe K$_\alpha$ line remains narrow, the Compton hump is weak, and all spectral fits point to low reflection fractions together with a warm corona component that dominates the soft X-ray emission. Porquet et al. (2018) showed that in the case of Ark 120, even though the source exhibits a broad Fe K$_{\alpha}$ line, the overall spectrum is best explained by invoking a warm, extended corona along with low reflection fraction. The low coherence (Figure 4) indicates that most of the soft X-ray excess varies incoherently with the hard band, similar to the behavior seen in PG 1211+143 (De Marco et al. 2011) and PG 1244+026 (Jin et al. 2013), where such low coherence ruled out a simple single reflection origin.
 Considering these circumstances, the classical reflection scenario becomes less persuasive, and alternative mechanisms such as lags within the warm Comptonizing region (e.g. Gardner \& Done 2014; Wilkins \& Gallo 2015; Middei et al. 2020) provide a more natural explanation for the modest soft lag observed in Mrk 110, if we consider the lower BH mass as discussed above. 
 Soft X-ray reverberation lags have been reported in a number of AGNs (e.g., De Marco et al. 2013). In several well studied sources, including 1H 0707-495, IRAS 13224-3809, Mrk 766, and MCG-6-30-15, the soft lags are most naturally explained by relativistic reflection, supported by the detection of Fe K$_{\alpha}$ reverberation and characteristic lag energy spectra (Fabian et al. 2009; Zoghbi et al. 2010; Kara et al. 2013, 2014; Emmanoulopoulos et al. 2014). But there are sources where soft lags are being reported following the scaling relations of De Marco et al. (2013) but deeper spectral studies revealed many other features like extended reprocessor along the disk eg. NGC 4051 (Turner et al. 2017), warm corona along with reflection in Ark 564 (Yu, Wilkins \& Allen 2025).

Table 3 displays few AGN sources exhibiting soft lags along with other parameters for comparison.
In the bare AGN Ark 120 ($M_{\rm BH}$ = 1.5 $\times$ 10$^{8}$ M$_{\odot}$, Peterson et al. 2004), soft lags of amplitude $\tau$ = $-$940$\pm$400 s were present at 5 $\times$ 10$^{-5}$ Hz and 8 $\times$ 10$^{-5}$ Hz (Lobban et al. 2018). The source displayed a low X-ray flux variability in 2013 (Matt et al. 2014) but in 2014 exhibited a high flux variability (Porquet et al. 2018). Mahmoud et al. (2023) found a warm corona temperature of $kT_{\rm w} \sim$ 0.34 -- 0.36 keV, slightly higher than that inferred for Mrk 110 assuming a $10^{8}$  M$_{\odot}$ black hole but consistent with the lower mass scenario (Table 2). From their study, the hot corona size remained stable ($R_{\rm h} \sim 23 R_{\rm g}$) while the warm corona expanded from $R_{\rm w} \sim $ 46 to $\sim$ 61 R$_{\rm g}$ between low and high flux states in Ark 120, trend similar to that seen in the Mrk 110 in the present study. AGN Mrk 509 displays a soft lag of $\sim$ $-$600 s (De Marco et al. 2013). It has a warm corona dominated spectrum (Petrucci et al. 2013), though reflection cannot be fully ruled out (García et al. 2019). Mrk 509 shows $kT_{\rm w}$  $\sim$ 0.2 keV obtained from AGNSED analysis (Kubota \& Done 2018). 
In case of AGN PG 1211+143 with a black hole of mass $M_{\rm BH}$ = 1.5 $\times$ 10$^{8}$ M$_{\odot}$ displayed a lag of $-$300 s -- $-$500 s in a frequency range of 10$^{-3}$ -- 10$^{-4}$ Hz. Later Lobban et al. (2018) reported a soft lag of $\le$ $-$800 s around 10$^{-4}$ Hz. Recently, Xu et al. (2024) observed a soft lag of $\sim$ $-$5000 s in a low flux state at 1.5 $\times$ 10$^{-5}$ Hz, which disappeared in a high flux state at the same frequency. 
In another source Mrk 704 with a black hole mass of $M_{\rm BH}$ = 1.3 $\times$ 10$^{8}$ M$_{\odot}$, Sriram et al. (2023) reported a soft lag $\sim$ $-$300 s in a frequency domain of 10$^{-4}$ -- 10$^{-3}$ Hz and argued that these soft lags were influenced by a warm absorber in the inner region of the accretion disk.

The seed photon temperature remained low in all observations (kT$_{\rm bb}$ = 7--9 eV) consistent with an origin in the outer accretion disk. The expected disk temperature can be estimated from the following relation (Netzer et al. 2013) 

\begin{equation}
kT_{\rm eff}(r) =
111~\,
\left(\frac{\dot{M}}{\dot{M}_{\rm Edd}}\right)^{1/4}
M_8^{-1/4}
\left(\frac{r}{r_g}\right)^{-3/4}
\left[1-\sqrt{\frac{r_{\rm in}}{r}}\right]^{1/4} eV
\end{equation}

where M$_{8}$ is the mass of the black hole in units of 10$^{8}$ M$_{\odot}$, r is the distance from the central black hole, $r_g = \frac{G M}{c^2}$, $\dot{M}/\dot{M}_{{Edd}}$ is the relative accretion rate. We calculated the peak disk temperatures observed at radius r $\sim$ 1.69 r$_{\rm g}$ for the Kerr black hole for accretion rate $\dot{M}/\dot{M}_{{Edd}}$ = 0.1. We noted that for a black hole mass of M $\sim$ 1.4 $\times$ 10$^{8}$ M$_{\odot}$, inner disk temperature is kT$_{\rm in}$ $\sim$ 24 eV and for M $\sim$ 2 $\times$ 10$^{7}$ M$_{\odot}$, kT$_{\rm in}$ $\sim$ 40 eV. Moreover the color correction factor would further increase the observed inner disk temperature. As a result, the relatively low seed photon temperature alone cannot be used to uniquely infer the black hole mass, leaving an ambiguity in the mass determination.


Hagen et al. (2024) showed that for $L/L_{\rm Edd} \gtrsim 0.02$ the accretion disk remains stable enough to sustain an extended warm corona, whereas at lower accretion rates the warm corona weakens as the disk collapses. 
The measured $L/L_{\rm Edd}$ = 0.023 -- 0.035 agrees with Porquet et al. (2024).
We noted that in the highest luminosity state (Obs. 2d) we find $R_{\rm h} = 15 \pm 2 R_{\rm g}$ and $R_{\rm w} = 183 \pm 38 R_{\rm g}$, compared to $R_{\rm h} = 13 \pm 3 R_{\rm g}$ and $R_{\rm w} = 80 \pm 24 R_{\rm g}$ in the lowest state (Obs. 2a), consistent with AGNSED predictions (Kubota \& Done 2018; Hagen et al. 2024) that $R_{\rm w}$ increases with $\dot{\rm m}$. However, we did not notice any smooth variation of R$_{\rm w}$ along with the luminosity. For a lower mass (M = 2$\times10^{7} M_{\odot}$), the spectral result gives $\log\dot{\rm m}$ = $-$0.43 to $-$0.70 which results in $L/L_{\rm Edd}$ = 0.20 -- 0.37 which is consistent with the results of Meyer-Hofmeister \& Meyer (2011) and noted a mild variation in R$_{\rm w}$ between Obs. 2d and Obs. 2a.

During our analysis, we found that even when the inner emissivity index varied over a broad range, (q$_{1}$ $\sim$ 3--10), the reflection fraction consistently remained low (R$_{\mathrm{f}}$ $<$ 1). A steep emissivity profile is usually accompanied by a strong reflection component if the inner accretion disk extends close to the ISCO and intercepts a large fraction of the coronal photons. The combination of high q$_{1}$ but low reflection fraction instead suggests a geometry in which only a small fraction of the emitted coronal radiation reaches the disk. This scenario is consistent with an anisotropic or extended corona, such as a mildly relativistic outflow or coronal jet, which beams a significant portion of its emission away from the disk (Beloborodov 1999; Wilkins \& Gallo 2015; Chainakun et al. 2016; Wilkins et al. 2022). In such configurations, gravitational light bending can still steepen the emissivity profile near the black hole, but the effective reflection strength is suppressed because the disk intercepts fewer photons. This scenario is further supported by the detection of relativistic jets in Mrk 110 during 2015-16 and 2021-2024 observations (Wang et al. 2025). In Mrk 335, the reflection fraction was found to be low during flare episodes, consistent with partial ejection of the coronal region (Wilkins \& Gallo 2015), and transient jet activity has also been detected (e.g. Yao et al. 2021). Similarly, Chamani et al. (2020) observed weak X-ray reflection during radio emitting phases in III Zw 2. Taken together, Mrk 110, Mrk 335, and III Zw 2 present a coherent picture in which episodic, low power, compact jets arise very close to the black hole and are associated with persistently low X-ray reflection fractions.}

\section{Conclusions}
Based on a comprehensive timing and spectral analysis of Mrk 110, we draw the following conclusions:\\

1. The modest soft lags detected in Mrk 110, with amplitudes $\sim$ 1000 s, are consistent with reverberation timescales expected from reflection in the inner accretion disk assuming a mass M = 1.4 $\times$ 10$^{8}$ M$_{\odot}$. The agreement between the frequency-lag, energy-lag and cross-correlation function analyses further supports a reflection based interpretation of the observed lags. Although the spectral modeling indicates the presence of a warm, optically thick corona, its contribution to the observed lag is likely secondary. 
However, assuming a black hole mass of M $\sim$ 2 $\times$ 10$^{7}$ M$_{\odot}$, the radius inferred from the observed soft lag suggests a contribution from both relativistic reflection and variability associated with the warm corona. However, given that the statistical significance of the detected lags is limited to the $\sim$ 80\% confidence level, and hence, will require confirmation with higher quality observations.

2. We attempted to explain the UV -- X-ray spectra all six observations with a double Comptonization model. We noted a low electron temperature $\sim$ 0.21 -- 0.24 keV along with high optical depth $\tau$ $\sim$ 15 in all observations irrespective of flux variability. The seed photon temperature of kT$_{\rm bb}$ $\sim$ 7 -- 9 eV favors a mass of M $\sim$ 1.4 $\times$ 10$^{8}$ M$_{\odot}$ for Mrk 110. A fractional soft flux variation of $\sim$ 26\% is observed but none of the spectral parameters of either the warm or the hot corona have changed.  \\

3. Assuming both the masses, we noted the warm corona with a $kT_{\rm w}$ $\le$ 0.33 keV along with a hot corona constrained to $R_{\rm h}$ $\sim$ 10 $R_{\rm g}$ and a warm corona extending in the range of $R_{\rm w}$ = 100 -- 200 $R_{\rm g}$ in Mrk 110. A larger warm corona radius (R$_{\rm w}$) is found to be associated with higher mass accretion rate, while a smaller 
R$_{\rm w}$ corresponds to lower accretion states; however, we do not find a continuous or monotonic trend of R$_{\rm w}$ across the full range of mass accretion rates. The high accretion rate ($\dot{m}$ $\sim$ 0.25 -- 0.40) obtained for the low mass SMBH is broadly consistent with previous studies (eg. Meyer-Hofmeister \& Meyer 2011; Du \& Wang 2019). Considering both masses, the reflection fraction remains low, supporting an outflowing corona geometry. This interpretation is further supported by the presence of radio jets reported in Mrk 110 (Wang et al. 2025), although no simultaneous radio data are available for the epochs analyzed in this work.

\section*{Acknowledgements}

We are grateful to the Referee for the insightful and constructive comments, which significantly improved the clarity and quality of this work.
K. S. and D. L. acknowledge the support from the ANRF CRG project, Government of India. VKA thanks GH, SAG; DD, PDMSA, and Director URSC for their encouragement and continuous support to carry out this research. This work is based on observations obtained with {\it XMM–Newton}, an ESA science mission with instruments and contributions directly funded by ESA Member States and NASA.

\onecolumn
\begin{table}[t]
\caption{Best fit for the observations using double Comptonization model with blackbody disk component(thcomp(w)*zbbody + thcomp(h)*zbbody+ZGauss).Flux$_{\rm Total}$ in the 0.3 -- 10 keV energy band with a units of $\times$ 10$^{-11}$ erg cm$^{-2}$ s$^{-1}$. Here subscript (w) and (h) denotes the parameters of the warm and hot corona. f is the fixed parameter at 30 keV.}
\begin{tabular}{cc cccc c}
\hline\hline
Parameters&2004 & \multicolumn{4}{c}{2019} & 2020 \\
\cline{3-6}
 & Obs. 1 & Obs. 2a & Obs. 2b & Obs. 2c & Obs. 2d & Obs. 3  \\
\hline
\hline
kT$_{e} (keV) $&0.22$^{+0.01}_{-0.01}$&0.33$^{+0.03}_{-0.02}$&0.34$^{+0.03}_{-0.02}$&0.33$^{+0.03}_{-0.03}$&0.32$^{+0.04}_{-0.03}$&0.29$^{+0.02}_{-0.02}$\\ 

$\Gamma$$_{w}$ (keV)&2.32$^{+0.09}_{-0.07}$&2.52$^{+0.04}_{-0.05}$&2.57$^{+0.03}_{-0.04}$&2.57$^{+0.04}_{-0.05}$&2.55$^{+0.04}_{-0.05}$&2.37$^{+0.04}_{-0.04}$\\

Norm$_{w}$$\times$10$^{-4}$&5.91$^{+0.15}_{-0.15}$&0.16$^{+0.11}_{-0.13}$&0.82$^{+0.33}_{-0.39}$&1.47$^{+0.26}_{-0.23}$&2.09$^{+0.48}_{-0.38}$&3.12$^{+0.34}_{-0.31}$\\

$kT_{\rm h}$ (keV)&30&f&f&f&f&f\\ 
$\Gamma$$_{h}$&1.77$^{+0.01}_{-0.01}$&1.67$^{+0.02}_{-0.02}$&1.70$^{+0.02}_{-0.02}$&1.73$^{+0.02}_{-0.02}$&1.67$^{+0.03}_{-0.03}$&1.67$^{+0.02}_{-0.02}$\\

 C$_{\rm f, h}$ ($\times$ 10$^{-3}$)&13$\pm$4&11$\pm$3&15$\pm$4&17$\pm$5&18$\pm$6&14$\pm$4\\
Norm$_{h}$$\times$10$^{-2}$&0.49$^{+0.02}_{-0.02}$&0.99$^{+0.04}_{-0.03}$&0.94$^{+0.07}_{-0.07}$&1.09$^{+0.04}_{-0.04}$&1.09$^{+0.06}_{-0.06}$&0.76$^{+0.04}_{-0.04}$\\
Flux$_{Total}$&6.30$\pm$0.01&4.16$\pm$0.01&5.66$\pm$0.01&6.08$\pm$0.01&6.45$\pm$0.01&5.00$\pm$0.01\\ 
$\chi^{2}$/dof&2716/2545&2454/2377&2815/2694&2644/2678&1944/2079&2382/2484\\ 
\hline
\hline
\end{tabular}
\end{table}

\onecolumn
\begin{table}
\centering
\caption{Best fit for all six observations using AGNSED + Relxillcp model. Here subscript (w) denotes the parameters of the warm corona and subscript (h) denotes the parameters of the hot corona. Few parameters are fixed at their respective values as shown in the column of Obs. 1.}
\begin{tabular}{cc cccc c}
\hline\hline
Parameters&2004 & \multicolumn{4}{c}{2019} & 2020 \\
\cline{3-6}
 & Obs. 1 & Obs. 2a & Obs. 2b & Obs. 2c & Obs. 2d & Obs. 3  \\
 
AGNSED&   && &&&\\
\hline
M = 1.4$\times$10$^{8}$ M$_{\odot}$&&&&&&\\
\hline
log $\dot{m}$ &-1.48$^{+0.}_{-0.}$&-1.60$^{+0.05}_{-0.02}$&-1.54$^{+0.03}_{-0.02}$&-1.55$^{+0.02}_{-0.03}$&-1.46$^{+0.02}_{-0.03}$&-1.59$^{+0.01}_{-0.02}$ \\

$kT_{\rm h}$ (keV)&30&f&f&f&f&f\\ 

kT$_{w} (keV) $&0.18$^{+0.02}_{-0.02}$&0.23$^{+0.02}_{-0.01}$&0.27$^{+0.01}_{-0.01}$&0.22$^{+0.01}_{-0.01}$&0.22$^{+0.01}_{-0.02}$&0.21$^{+0.02}_{-0.01}$\\ 

$\Gamma$$_{h}$&1.83$^{+0.01}_{-0.02}$&1.73$^{+0.01}_{-0.02}$&1.80$^{+0.01}_{-0.02}$&1.83$^{+0.01}_{-0.01}$&1.91$^{+0.05}_{-0.06}$&1.76$^{+0.01}_{-0.01}$ \\

$\Gamma$$_{w}$&2.20$^{+0.03}_{-0.04}$&2.30$^{+0.02}_{-0.04}$&2.39$^{+0.02}_{-0.03}$&2.32$^{+0.13}_{-0.14}$&2.34$^{+0.02}_{-0.01}$&2.14$^{+0.02}_{-0.02}$ \\

$R_{\rm h}$ ($R_{\rm g}$)&11$^{+2}_{-1}$&13$^{+2}_{-3}$&9$^{+3}_{-1}$&7$^{+2}_{-3}$&15$^{+1}_{-2}$&13$^{+4}_{-5}$ \\

$R_{\rm w}$ ($R_{\rm g}$)&105$^{+13}_{-10}$&81$^{+22}_{-24}$&100$^{+25}_{-28}$&97$^{+21}_{-20}$&183$^{+38}_{-36}$&82$^{+21}_{-22}$ \\

log R$_{out}$ ($R_{\rm g}$)&-1&f&f&f&f&f\\
h ($R_{\rm g}$) &10&f&f&f&f&f\\
Log R$_{out}$($R_{\rm g}$)&-1&f&f&f&f&f\\
cos i&0.9848&f&f&f&f&f\\
Reprocess&1&f&f&f&f&f\\

\hline
RelxillCp&&&&&&\\
\hline
R$_{\rm f}$&0.21$^{+0.03}_{-0.03}$&0.48$^{+0.15}_{-0.16}$&0.49$^{+0.06}_{-0.05}$&0.18$^{+0.05}_{-0.05}$&0.74$^{+0.05}_{-0.05}$&0.43$^{+0.03}_{-0.03}$\\
Norm$\times$10$^{-4}$&6.44$^{+1.28}_{-2.23}$&2.84$^{+0.023}_{-0.022}$&4.61$^{+0.034}_{-0.033}$&5.04$^{+0.021}_{-0.023}$&2.41$^{+0.005}_{-0.006}$&7.51$^{+0.067}_{-0.065}$\\

\hline

$\chi^{2}$/dof&2710/2551&2557/2382&2945/2699&2879/2682&1980/2084&2353/2487\\ 

\hline
\hline
M = 2$\times$ 10$^{7}$ M$_{\odot}$ &&&&&&\\
\hline
\hline
log $\dot{m}$ &-0.50$^{+0.03}_{-0.02}$&-0.70$^{+0.02}_{-0.01}$&-0.56$^{+0.02}_{-0.03}$&-0.55$^{+0.01}_{-0.01}$&-0.43$^{+0.01}_{-0.02}$&-0.54$^{+0.01}_{-0.01}$ \\

$kT_{\rm h}$ (keV)&30&f&f&f&f&f\\ 

kT$_{w} (keV) $&0.27$^{+0.01}_{-0.02}$&0.32$^{+0.02}_{-0.03}$&0.31$^{+0.02}_{-0.03}$&0.33$^{+0.03}_{-0.03}$&0.33$^{+0.23}_{-0.23}$&0.31$^{+0.02}_{-0.01}$\\ 

$\Gamma_{h}$&1.84$^{+0.01}_{-0.02}$&1.70$^{+0.02}_{-0.02}$&1.78$^{+0.02}_{-0.01}$&1.79$^{+0.01}_{-0.02}$&1.76$^{+0.02}_{-0.02}$&1.72$^{+0.02}_{-0.03}$ \\

$\Gamma_{w}$&2.57$^{+0.06}_{-0.07}$&2.53$^{+0.04}_{-0.05}$&2.55$^{+0.03}_{-0.03}$&2.61$^{+0.03}_{-0.04}$&2.59$^{+0.02}_{-0.01}$&2.41$^{+0.04}_{-0.05}$ \\

$R_{\rm h}$ ($R_{\rm g}$)&13$^{+3}_{-2}$&11$^{+2}_{-3}$&10$^{+3}_{-1}$&12$^{+2}_{-3}$&22$^{+1}_{-2}$&12$^{+4}_{-5}$ \\

$R_{\rm w}$ ($R_{\rm g}$)&198$^{+10}_{-19}$&160$^{+14}_{-19}$&190$^{+11}_{-16}$&184$^{+13}_{-11}$&213$^{+15}_{-14}$&216$^{+11}_{-10}$ \\


\hline
RelxillCp&&&&&&\\
\hline
R$_{\rm f}$&0.56$^{+0.06}_{-0.08}$&0.43$^{+0.06}_{-0.07}$&0.46$^{+0.04}_{-0.02}$&0.40$^{+0.04}_{-0.05}$&0.82$^{+0.23}_{-0.22}$&0.42$^{+0.05}_{-0.06}$\\

Norm $\times$10$^{-4}$&4.62$^{+0.05}_{-0.05}$&4.40$^{+0.03}_{-0.03}$&7.31$^{+1.22}_{-1.25}$&3.84$^{+0.53}_{-0.55}$&2.06$^{+0.11}_{-0.12}$&4.34$^{+0.32}_{-0.34}$\\

\hline

$\chi^{2}$/dof&2704/2551&2483/2382&2778/2699&2713/2682&1934/2084&2239/2487\\ 

\hline
\hline
\end{tabular}
\end{table}
\clearpage

\onecolumn
\begin{table}
\small
\caption{Log of a few AGNs exhibiting soft lags and warm corona properties. $kT_{\rm w}$ is the warm corona temperature. $R_{\rm h}$ and $R_{\rm w}$ determined from AGNSED model for few AGN sources.}    
\begin{tabular}{c c c c c c c}
\hline   
 Source &Frequency  &Soft Lag& $kT_{\rm w}$ & $R_{\rm h}$,  $R_{\rm w}$  &SMBH Mass &References\\

 & (Hz)& (s)&(keV)&($R_{\rm g}$)&(10$^{8}$ M$_{\odot}$)&\\
  \hline
 Mrk 110 &7-9 $\times$ 10$^{-5}$ & $-$889 -- $-$3067  &0.20 -- 0.25 & 10 -- 20 , 100 -- 200 &0.2, 1.4 &Present work\\ \\

 Ark 120 &5$\times$10$^{-5}$ --  8$\times$10$^{-5}$& -940$\pm$400 &0.36& 23, 46-61  &1.5 &Lobban et al. (2018) \\
 &&&&&&Mahmoud et al. (2023)\\

 
 Mrk 509&10$^{-4}$ -- 10$^{-5}$& $\sim$ $-$600 &0.4-0.5 & &1.3 & Petrucci et al. (2013)\\ 
 & & &0.2 & 21 / 40 & 1.0 &Kubota \& Done (2018)\\ \\
 PG1211+143&10$^{-4}$&$\le$ $-$800 & --& --& 1.4  &Lobban et al. (2018)\\
 & 1.5 $\times$ 10$^{-5}$ & $\sim$ $-$5000 &--&--&-- & Xu et al. (2024)\\ \\
 Mrk 704&10$^{-3}$ -- 10$^{-4}$&$\sim$ $-$300&--&- &1.3&Sriram et al. (2023) \\ \\

\hline
\end{tabular}
\end{table}

\newpage
{}

\end{document}